%% file: sample7.tex
\documentclass[trackchanges, twocolumn]{aastex7}

\newcommand\tp{\omega(\theta)}
\newcommand\lcorr{\l_{\mathrm{corr}}}

\usepackage{amsmath}
\usepackage{tabularx}
\usepackage{makecell}
\usepackage{hyperref}
\received{August 29th, 2025}
\revised{November 20th, 2025}
\accepted{January 15th, 2026}

\begin{document}

\title{FEAST: Probing Hierarchical Star Formation with the Spatial Distributions of Young Star Clusters}

\defcitealias{menon21}{M21}

\correspondingauthor{Drew Lapeer}
\email{dlapeer@umass.edu}

\author[0009-0009-5509-4706]{Drew Lapeer}
\affiliation{Department of Astronomy, University of Massachusetts, 710 North Pleasant Street, Amherst, MA 01003, USA}
\email{dlapeer@umass.edu}

\author[0000-0002-5189-8004]{Daniela Calzetti}
\affiliation{Department of Astronomy, University of Massachusetts, 710 North Pleasant Street, Amherst, MA 01003, USA}
\email{dcalzetti@umass.edu}

\author[0000-0002-3247-5321]{Kathryn~Grasha}
\altaffiliation{ARC DECRA Fellow}
\affiliation{Research School of Astronomy and Astrophysics, Australian National University, Canberra, ACT 2611, Australia}  
\email{Kathryn.Grasha@anu.edu.au}

\author[0000-0002-8192-8091]{Angela Adamo}
\affiliation{Department of Astronomy, Oskar Klein Centre, Stockholm University, AlbaNova University Center, SE-106 91 Stockholm, Sweden}
\email{angela.adamo@astro.su.se}

\author[0000-0002-1723-6330]{Bruce G. Elmegreen}
\affiliation{Katonah, New York USA 10536}
\email{belmegreen@gmail.com}

\author[0000-0001-8068-0891]{Arjan Bik}
\affiliation{Department of Astronomy, Oskar Klein Centre, Stockholm University, AlbaNova University Center, SE-106 91 Stockholm, Sweden}
\email{arjan.bik@astro.su.se}

\author[0009-0003-6182-8928]{Giacomo Bortolini}
\affiliation{Department of Astronomy, The Oskar Klein Centre, Stockholm University, AlbaNova, SE-10691 Stockholm, Sweden}
\email{giacomo.bortolini@astro.su.se}

\author[0000-0002-1832-8229]{Anne Buckner}
\affiliation{Cardiff Hub for Astrophysics Research and Technology (CHART), School of Physics \& Astronomy, Cardiff University, The Parade, CF24 3AA Cardiff, UK}
\email{BucknerA@cardiff.ac.uk}

\author[0000-0001-6291-6813]{Michele Cignoni}
\affiliation{Dipartimento di Fisica, Università di Pisa, Largo Bruno Pontecorvo 3, 56127, Pisa, Italy } 
\affiliation{INFN, Largo B. Pontecorvo 3, 56127, Pisa, Italy} 
\affiliation{INAF - Osservatorio di Astrofisica e Scienza dello Spazio di Bologna, Via Piero Gobetti 93/3, 40129, Bologna, Italy}
\email{michele.cignoni@unipi.it}

\author[0000-0001-6464-3257]{Matteo Correnti}
\affiliation{INAF Osservatorio Astronomico di Roma, Via Frascati 33, 00078, Monteporzio Catone, Rome, Italy}
\affiliation{ASI-Space Science Data Center, Via del Politecnico, I-00133, Rome, Italy}
\email{matteo.correnti@inaf.it}

\author[0000-0002-1392-3520]{Debra Meloy Elmegreen}
\affiliation{Vassar College, Dept. of Physics \& Astronomy, Poughkeepsie, New York USA 12604}
\email{elmegreen@vassar.edu}

 \author[0000-0002-2199-0977]{H. Faustino Vieira}
\affiliation{Department of Astronomy, Oskar Klein Centre, Stockholm University, AlbaNova University Centre, SE-106 91 Stockholm, Sweden}
\email{helena.faustinovieira@astro.su.se}

\author[0009-0001-2656-4299]{Max Hamilton}
\affiliation{Manning College of Information and Computer Sciences, 140 Governors Dr, Amherst, MA 01003, USA}
\email{jmhamilton@umass.edu}

\author[0000-0001-8348-2671]{Kelsey Johnson}
\affiliation{
Department of Astronomy, University of Virginia, Charlottesville, VA 22904, USA
}
\email{kej7a@virginia.edu}

\author[0000-0001-8490-6632]{Thomas S.-Y. Lai}
\affil{IPAC, California Institute of Technology, 1200 E. California Blvd., Pasadena, CA 91125, USA}
\email{ThomasLai.astro@gmail.com}

\author[0000-0002-1000-6081]{Sean T. Linden}
\affiliation{Steward Observatory, University of Arizona, 933 N Cherry Avenue, Tucson, AZ 85721, USA}
\email{seanlinden@arizona.edu}

\author[0000-0002-3869-9334]{Subhransu Maji}
\affiliation{Manning College of Information and Computer Sciences, 140 Governors Dr, Amherst, MA 01003, USA}
\email{smaji@umass.edu}

\author[0000-0003-1427-2456]{Matteo Messa}
\affiliation{INAF - Osservatorio di Astrofisica e Scienza dello Spazio di Bologna, Via Gobetti 93/3, I-40129 Bologna, Italy}
\email{matteo.messa@inaf.it}

\author[0000-0002-3005-1349]{Göran Östlin}
\affiliation{ Department of Astronomy, Stockholm University, Oscar Klein Centre, AlbaNova University Center, 10691 Stockholm, Sweden}
\email{ostlin@astro.su.se}

\author[0000-0002-8222-8986]{Alex Pedrini}
\affiliation{Department of Astronomy, Oskar Klein center, Stockholm University, AlbaNova University center, SE-106 91 Stockholm, Sweden}
\email{alex.pedrini@astro.su.se}

\author[0000-0003-2954-7643]{E. Sabbi}
\affiliation{Gemini Observatory/NSFs NOIRLab, 950 N. Cherry Ave., Tucson, AZ 85719, USA}
\email{elena.sabbi@noirlab.edu}

\author[0000-0002-0806-168X]{Linda J. Smith}
\affiliation{Space Telescope Science Institute, 3700 San Martin Drive, Baltimore, MD 21218, USA}
\email{lsmith@stsci.edu}

\begin{abstract}
We apply the angular two-point correlation function (TPCF) to the spatial distribution of young star clusters (YSCs) in four nearby star forming galaxies (NGC 628, NGC 4449, M51, and M83) in order to investigate their underlying hierarchical structuring. Using newly constructed catalogs of YSCs in the emerging phase (eYSCs), identified in the infrared with JWST, and optical YSCs detected in archival HST data, we compute TPCFs for various cluster samples and age bins across the four galaxies as part of the FEAST (Feedback in Emerging extrAgalactic Star ClusTers) program. We find clear evidence of hierarchical structuring, especially in eYSCs and YSCs with ages $<10$ Myr (referred to as oYSCs), which show similar TPCFs within each galaxy. NGC 628 exhibits a clear distinction between the TPCFs of eYSCs and oYSCs, implying a shorter randomization timescale. In contrast, clusters aged 10–300 Myr exhibit progressively more random spatial distributions, becoming effectively random after $\sim 100$ Myr, consistent with earlier studies. The two-dimensional fractal index $D_2$ of the YSCs underlying distribution is calculated from model fits to TPCFs. Our values of $D_2$ derived from the youngest YSC populations align better with the expected value of $D_2 \sim 1.3$ for a universal star formation process compared to previous findings.

\end{abstract}


\section{Introduction}\label{sec:intro}

It is well established that the majority of stars form in some sort of stellar aggregate \citep{ladalada03}. The densest types of stellar aggregate are star clusters: gravitationally bound stellar systems with radii of order 0.5--10 parsecs and masses $10^3 < M_*/M_{\odot} < 10^7$  \citep{pz10, ryon17, brown21}. Star clusters are direct products of the star formation process, and inherit the fractal structure of the interstellar medium (ISM) they are born within \citep{egsalzar99, elmegreen03, elmegreenscalo04, elmegreen07, bergin2007, beattie19, krumholz19}. Their bright nature and intimate link to the ISM makes young star clusters (YSCs) a fruitful avenue for probing the physical processes governing hierarchical star formation.

The star-formation process is hierarchical and self-similar in both space and time \citep{scalo85, li2005, elmegreen10, grasha17}, ranging from kpc to pc scales. The scale-free, fractal-like nature of the ISM is thought to be set by turbulence and gravitation \citep{elmegreen89}, both of which are key-processes governing current theories of star formation \citep{elmegreen_efremov96}. In this hierarchy, dense regions are nestled within less dense regions, with stars and star clusters forming in the densest regions of the hierarchy. The majority of regions in this hierarchy are unbound. YSCs inherit the structure of this hierarchy from their natal clouds \citep{grasha18, turner22} and remain bound for several-to-tens of Myr. This, paired with their bright (observable) nature, makes YSCs useful tracers of underlying hierarchical structure and allows for comparison across a range of ages. This makes their spatial distribution especially useful in tracing the underlying ISM structure. One way to gain insight from this spatial distribution is the two-point correlation function (TPCF). The TPCF quantifies the degree of clustering in a distribution of points through comparison to a similar, random distribution at a given scale \citep{peebles80}. Indeed, efforts have been made to investigate the hierarchical star formation process using the spatial distributions of YSCs in local volume galaxies \citep{zhang2001, bastian05, grasha_15, grasha17, menon21}. 

The spatial resolution and sensitivity of HST provided the ability to obtain large catalogs of star clusters \citep[e.g.,][]{whitmore11, chandar14, johnson16, legus_clusters, messa18} out to $\lesssim 20-30$ Mpc, and much of the previous work using the TPCF to probe hierarchical structure with YSCs used these catalogs \citep[e.g.,][]{grasha_15, grasha17, menon21, turner22}. However, the earliest stages of a YSC's life are obscured by their natal clouds and undetectable in the optical regime \citep{kobulnicky99, johnson01}. Consequently, previous surveys missed on average 60 per cent of YSCs in their embedded phase \citep{messa21}. The advent of JWST now allows us to probe this embedded phase, producing catalogs of embedded YSCs (eYSCs) \citep[e.g.,][A. Adamo et al 2025, in prep.]{whitmore23, rodriguez23, levy24, linden23, linden24, pedrini24, knutas25}. 

eYSCs are inherently more closely coupled to the ISM than clusters detected in the optical, making them prime candidates for probing the hierarchical and scale-free structure of the ISM through their spatial distributions. Using new eYSC catalogs and updated optical YSC catalogs produced by the Feedback in Emerging extrAgalactic Star ClusTers (FEAST, JWST GO 1783, PI A. Adamo) program, we extend previous work done using the TPCF of YSCs \citep{grasha_15, grasha17, menon21} into this new regime. 

This work is organized as follows. In Section \ref{sec:data}, we outline the sample of galaxies and the data used in this work. In Section \ref{sec:methods}, we outline our methods, providing a description of the TPCF and models used to characterize its features. Section \ref{sec:results} comprises our results and discussion. Finally, we conclude and summarize our work in Section \ref{sec:conclusions}.

\section{Data}\label{sec:data}

This work makes use of eYSC and YSC catalogs obtained with data covering four local volume, star forming galaxies from the JWST-FEAST program. In the sections that  follow, we describe our galaxy sample, and the star cluster catalogs used in this work. All our data products are available at MAST as a High Level Science Product via \dataset[10.17909/6dc1-9h53]{http://dx.doi.org/10.17909/6dc1-9h53} and on the FEAST webpage \href{https://feast-survey.github.io/}{https://feast-survey.github.io/}.

\subsection{Galaxy Sample}\label{sec:galaxies}

All 4 galaxies in this sample are local-volume ($D < 11$ Mpc) star-forming galaxies: NGCs~628, 4449, 5194, and 5236. They are all nearly face-on, grand design spiral galaxies, with the exception of NGC~4449, which is an irregular Magellanic type dwarf. A summary of the galaxies and their basic properties can be found in Table \ref{tab:galaxies}

\input{Table1}

\begin{table*}
\centering
\caption{HST data, science aperture radius, and sky annulus radius used to create the optical catalogs. HST/WFC3 data are from program numbers \#13340 (Van Dyk), \#12762 (Kuntz), \#13364 (Calzetti), \#17225 (Calzetti), \#11360 (O’Connell), and \#12513 (Blair). HST/ACS data are from program numbers \#10452 (Beckwith), \#9796 (Miller), \#10402 (Chandar), and \#10585 (Aloisi). All sky annuli have a width of 2 px.}
\label{tab:table2}
\begin{tabular}{lllll}
\hline
\hline
Galaxy & \makecell{HST/WFC3} & \makecell{HST/ACS} & \makecell{Aperture \\ Radius (px)} & \makecell{Sky Annulus \\ Radius (px)} \\
\hline
NGC 628 & \makecell{F275W, F336W} & \makecell{F435W, F555W, \\ F658N, F814W}& \makecell{4}& \makecell{6}\\
\hline
NGC 4449  & \makecell{F275W, F336W}& \makecell{F435W, F555W, \\ F658N, F814W} & \makecell{5} & \makecell{7}\\
\hline
M51  & \makecell{F275W, F336W, \\ F689M} & \makecell{F435W, F555W, \\ F658N, F814W} & \makecell{4} & \makecell{6} \\
\hline
M83 & \makecell{F225W, F275W, F336W, \\ F438W, F547M, F555W, \\ F657N, F689M, F814W} & \makecell{N/A} & \makecell{5}& \makecell{7} \\
\hline
\end{tabular}

\vspace{0.3cm}
\small
\end{table*}

\subsubsection{NGC 628}
NGC 628 is a grand design spiral galaxy (SAc), with a nearly face-on orientation. Located at a distance of 9.84 Mpc \citep{leroy_21}, it is the most distant galaxy in this sample. The hierarchical distribution of young star clusters in NGC 628 has been studied in the past \citep[][]{grasha_15, grasha17, menon21}. Previous work has used YSCs detected in the UV-optical from HST observations as part of the `Legacy ExtraGalactic UV Survey' (LEGUS) program \citep{legus_main, legus_clusters}. 

\subsubsection{NGC 4449}
NGC 4449 is an irregular Magellanic-like galaxy located at a distance of 4 Mpc \citep{legus_main}. It has an inclination of $68^{\circ}$ \citep{hunter05}. Both \cite{grasha17} and \cite{menon21} have performed analysis on the spatial distribution of YSCs in NGC 4449. It is the only dwarf galaxy in this sample. 

\subsubsection{M51}
M51 (NGC 5194) is a grand design spiral galaxy (SAbc) viewed at a nearly face-on orientation and located at a distance of 7.5 Mpc \citep{m51dist}. Unlike the other galaxies in this sample, NGC 5194 is currently undergoing an interaction with a companion galaxy (NGC 5195). Like the previous two galaxies, the hierarchical distribution of young star clusters in NGC 5194 has also been investigated in previous work \citep{grasha19, menon21}.

\subsubsection{M83}
M83 (NGC 5236) is a grand-design spiral galaxy (SABc) with a central bar. It is oriented nearly face-on and located at a distance of 4.7 Mpc \citep{dellabruna2022}. No previous investigations of this kind have been performed on M83.

\subsection{Star Cluster Catalogs}

In this study, we use the eYSC catalogs obtained from JWST NIRcam/MIRI observations of NGCs 628, 4449, 5194, and 5236 as part of the FEAST JWST program. Likewise, we use newly-derived YSC catalogs from archival HST imaging of the 4 galaxies as part of the FEAST program. eYSCs and YSC spectral energy distributions (SEDs) have been analyzed using CIGALE \citep{cigale} to derive the star cluster physical properties (ages, mass, extinction). In the optical YSC catalogues we select the sub-sample population of clusters with ages $<$10 Myr, here referred to as oYSCs. In the section that follows, we briefly describe these catalogs and the methods used to construct them. 

\subsubsection{Emerging Young Star Cluster Catalogs}
\label{sec:eyscclass}

Catalogs of eYSCs have been obtained from the \textit{JWST} NIRCam observations in 8 bands: F115W, F150W, F187N, F200W (NIRCam Short Wavelength channels) and F277W/F300M, F335M, F405N, F444W (NIRCam Long Wavelength channels). The construction of these catalogs and an in-depth description of the FEAST data are presented in \cite{gregg24} and \cite{knutas25}, and fully described in A. Adamo et al. 2025 (prep). 

Images are re-sampled to a scale of $0.04^{"}$/pix to match archival \textit{HST} data. From these data, emission maps of Pa$\alpha$, $4.05 \;\mu$m Br$\alpha$, and 3.3$\mu$m polycyclic aromatic hydrocarbon (PAH) are derived \citep[see][]{gregg24, pedrini24}. 

To extract the eYSCs, SourceExtractor \citep{sextractor} is run, and compact sources in the emission line maps are selected. This catalog is then visually inspected to discard any residuals or contaminants. Photometry is performed with a science aperture of radius 5px, and a sky annulus 2px wide with outer radius 7 px on all visually confirmed sources. An overlap criterion of 4 px is then applied. Detailed description of the eYSC catalogs of the FEAST galaxies can be found in Adamo et al (in prep.), \citet{knutas25}, and \citet{pedrini25b}.

eYSCs detected in these emission maps are divided into two categories.  Sources showing only recombination line emission (without 3.3 $\mu$m PAH emission) could be indicative of a spatial separation between the photo-dissociation and H II regions, while sources with co-spatial ionized H and 3.3 $\mu$m PAH emission could be more embedded in their natal H II region. In this framework, the former could be later evolutionary stages of the latter. Therefore, sources showing co-spatial peaked emission in ionized H and 3.3 $\mu$m PAH are denoted eYSC I, while sources showing only ionized H are denoted eYSC II.

\subsubsection{Optical Young Star Cluster Catalogs}
\label{sec:oyscClass}

In this work, we make use of  new optical YSC catalogs built from archival HST data and combined with JWST photometry as part of the FEAST program. Here we provide a brief description of the process used to create the catalogs. An overview of the photometry used to build the optical catalogs is given in Table \ref{tab:table2}. A careful description of the process is presented in \cite{knutas25} for the optical catalog of M83 and A. Adamo et al. 2025 (in prep.) for NGC 628. NGC4449 and M51 optical YSC catalog's construction is presented in \cite{pedrini25b}. 

For NGC4449 \citep{whitmore2020} and M51 \citep{messa18}, we used as starting point the cluster candidate positions and morphological classifications as delivered by the LEGUS project to rebuild the FEAST catalogues. Photometry was performed from the UV to the 5 $\mu$m using all HST and JWST available data with the in-house FEAST photometry pipeline. Photometry was extracted with a science aperture of radius 5 px (4 px for M51), and a sky annulus 2 px wide with outer radius 7 px (6 px for M51). Different aperture sizes are designed to preserve a similar physical scale across the sample.

In the case of NGC628, a new optical catalog was constructed because the LEGUS FoV did not sufficiently cover the area sampled by the JWST mosaic. For this catalogue, we performed source extraction in the HST F555W to obtain sources $>5\sigma$ above background levels. Sources with a concentration index (CI; the difference between F555W magnitude measured in a 1 pixel aperture and 3 pixel aperture, see \cite{legus_clusters}) broader than the stellar one (CI $\gtrsim$ 1.2,  \cite{knutas25})  were classified by STARCNET \citep{perez2021}, a machine-learning algorithm for automated star cluster classification. The final YSC catalogs include sources detected in the previous catalog \citep{legus_clusters} and newly detected objects. Photometry was performed from the UV to the 5 $\mu$m using all HST and JWST available data with the in-house FEAST photometry pipeline (see A. Adamo et al. 2026 in prep). Photometry was extracted with a science aperture of radius 4 px, and a sky annulus 2 px wide with radius 6 px. 

Finally, for M83 detections were made using  publicly available HST data in F555W and F547M which cover the central region and disk. This initial sample was culled to retain only sources with a signal-to-noise better than 5 in F438W, F555W or F547M, and F814W, absolute magnitude of at least -6 ABmag in F555W or F547M, and $\mathrm{CI} > 1.2$ mag to exclude single stars. Photometry was performed in 8 HST bands (F225W, F275W, F336W, F438W, F547M, F555W, F657N, F814W) and 8 NIRcam bands with radius of 5 pixels and a local sky annulus 2 pixels wide at 7 pixels. This catalog was then visually inspected and compared to the YSC catalog from \cite{dellabruna2022}. Our updated catalog includes 7419 of the 7459 sources from \cite{dellabruna2022} and adds an additional 455, resulting in 7874 total sources.

\subsubsection{Catalog Cuts}\label{sec:cuts}

We make several cuts to the star cluster catalogs before performing our analysis. As the YSC and eYSC catalogs are independently derived, we perform a cross-match between the two catalogs (for each galaxy) to make sure there is no cross contamination between eYSCs and YSCs. To do so, we use a circular aperture with radius 4 pixels that is centered on a YSC. If any eYSCs fall within the aperture, it is removed from the eYSC catalog (i.e., it is designated as an oYSC). Only $\sim 5$ \% or less of eYSCs are overlapping with an oYSC in a given galaxy. To ensure that the eYSCs are well-detected, we make a further photometric cut requiring $\mathrm{S/N} > 3$ in every available NIRCam band. In the optical catalogues only YSCs with morphological class 1 (compact, symmetric sources), 2 (compact, asymmetric sources), and 3 (multiple peak sources with diffuse extended emission) \citep{legus_clusters} are used in analysis. We also ensure that only sources within the JWST footprint are included in the analysis.

Finally, we make cuts based on the quality of a YSC's SED fit. For the eYSC and YSC catalog, we require that the CIGALE output results have a reduced $\chi^2 \leq 50$, following \cite{pedrini25b}. We likewise remove all clusters with ages $> 300$ Myr, and masses $< 200\;M_{\odot}$. The catalogs are complete down to $\sim 10^3\ M_{\odot}$ \citep[][]{pedrini25b, knutas25}.

\begin{figure*}
    \centering
    \includegraphics[width=\linewidth]{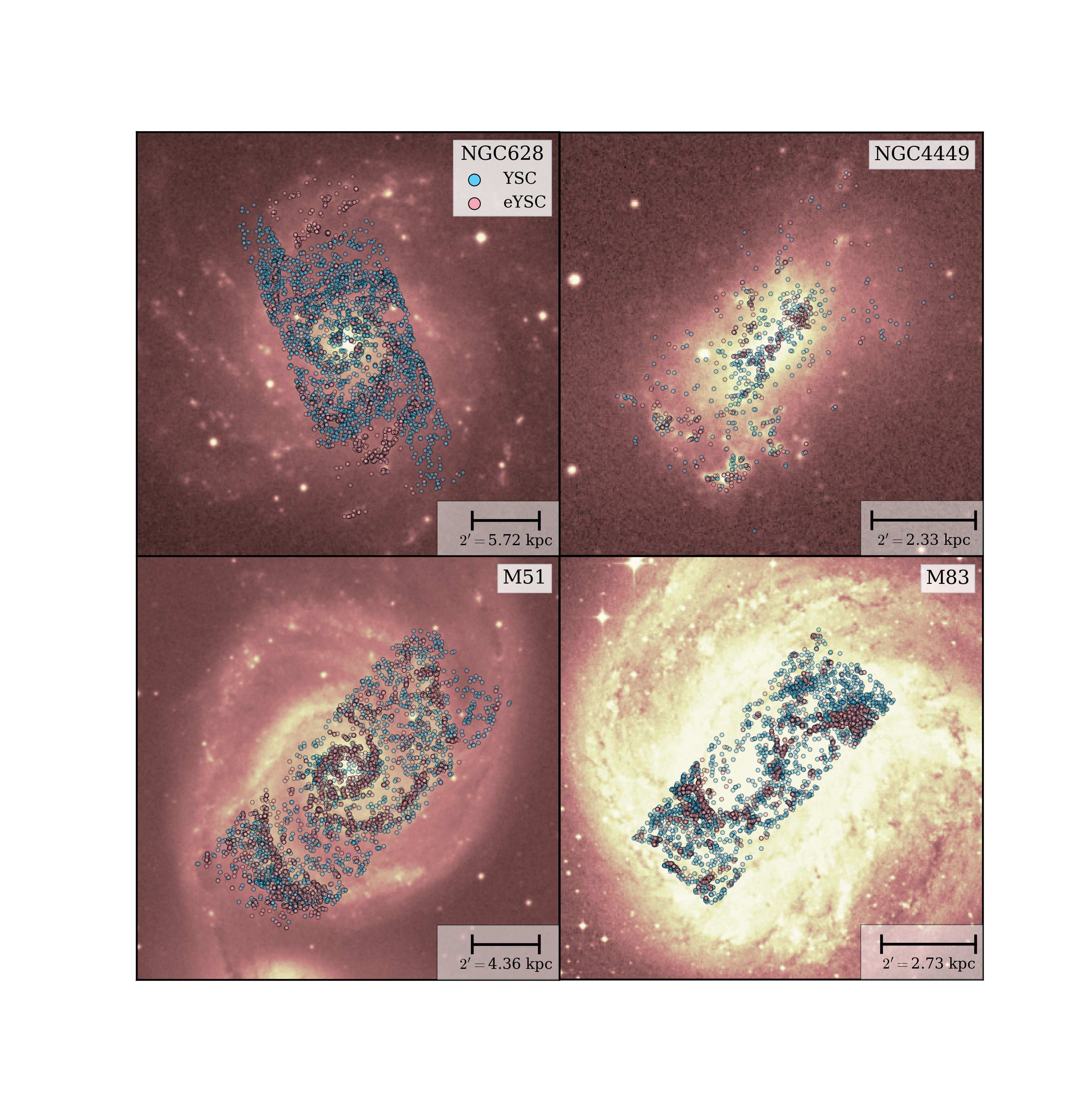}
    \caption{The positions of eYSCs (pink) and YSCs (blue) in each of this sample galaxies, with scale bars indicating length scales corresponding to $2^{'}$. Visually, eYSCs tend to trace spiral arm structure compared to YSCs, which are more evenly distributed throughout the galaxy. Background images of galaxies are taken from the Digitized Sky Survey \citep{dssref}.}
    \label{fig:galaxyImages}
\end{figure*}

\begin{figure}
    \centering
    \includegraphics[width=\linewidth]{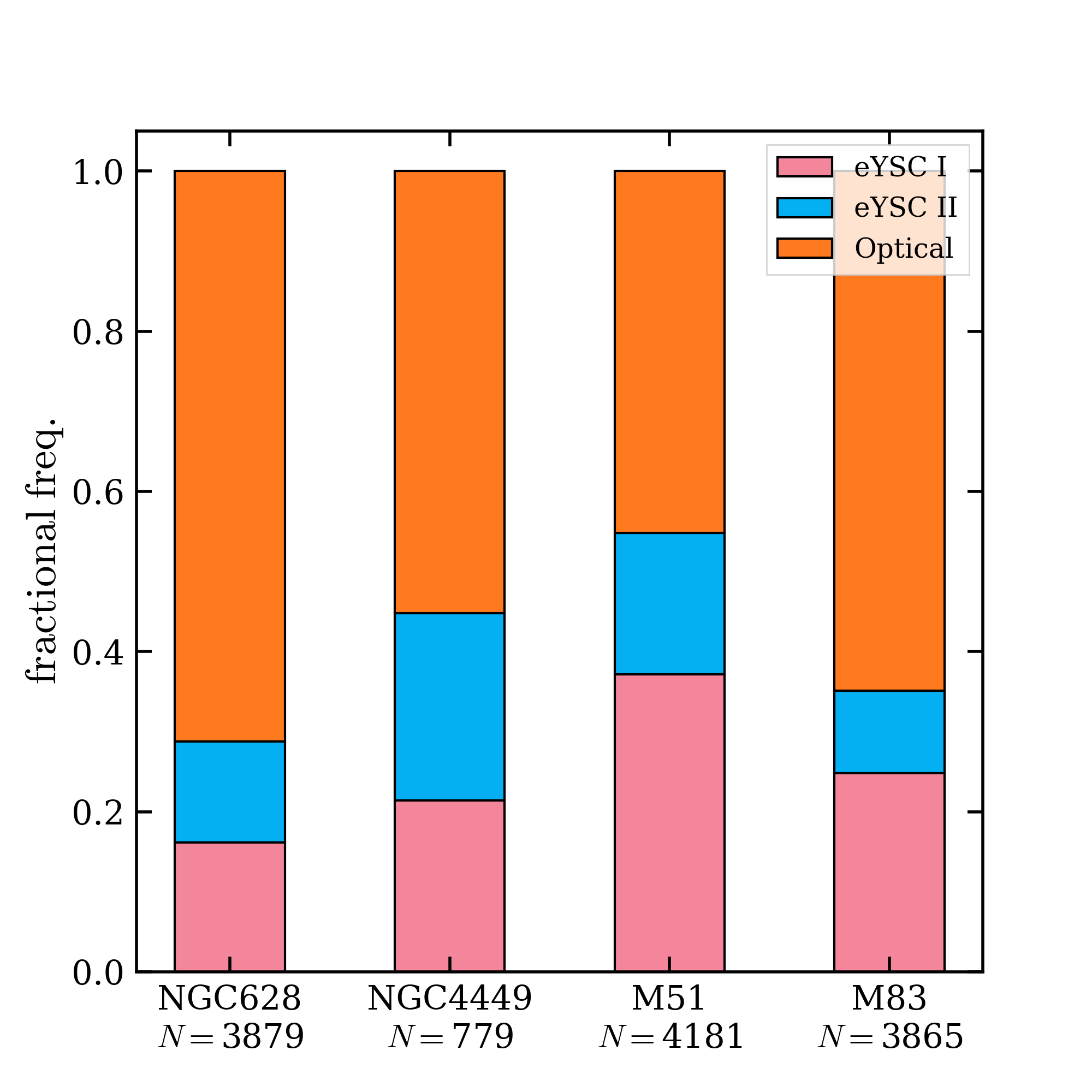}
    \caption{The relative frequencies of eYSC Is (orange), eYSC IIs (pink), and YSCs (blue) for each of our galaxies, designated on the $x$-axis. The total number of clusters is shown under the galaxy name. eYSCs tend to comprise $\sim 30-50$ \% of the post-cut catalog}
    \label{fig:clusterDist}
\end{figure}

Figure \ref{fig:galaxyImages} shows our samples overplotted over images of each galaxy in this sample. The final number of eYSC I and IIs, and YSCs are presented in Table \ref{tab:galaxies} and their respective fractions are shown in Figure \ref{fig:clusterDist} for each of our galaxies. The number of clusters used here varies slightly from what is presented in \cite{pedrini25b} due to our additional $\chi^2$ cut. We plot the CIGALE masses against ages of the YSCs in Figure \ref{fig:massAgeScatter} for each galaxy in this sample.  We address the possible impact of incompleteness in Appendix \ref{sec:completeness}, finding that the incompleteness has no major impact on our results and analysis.

\begin{figure*}
    \centering
    \includegraphics[width=\linewidth]{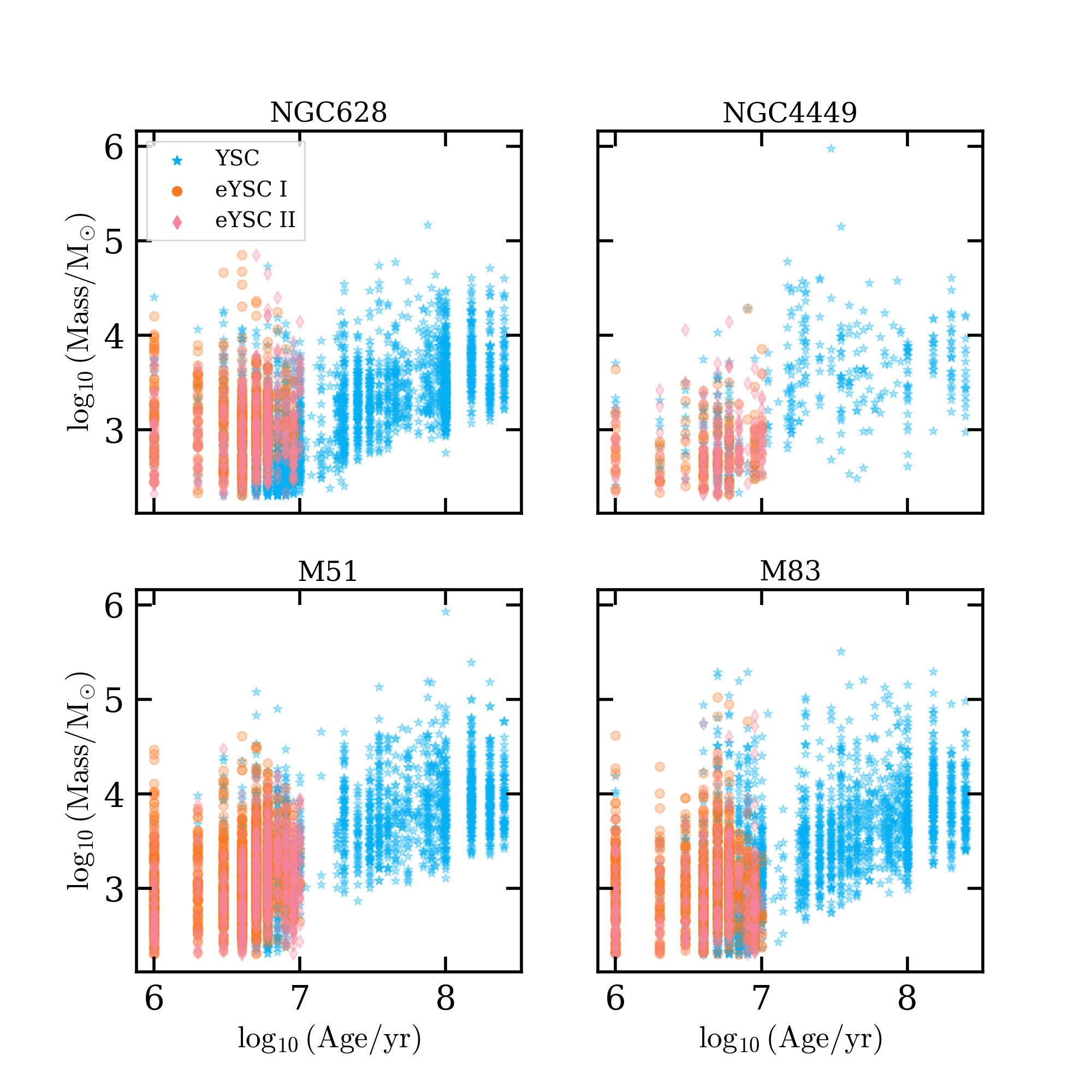}
    \caption{Mass plotted against age for the YSCs in each of the galaxies in this sample. Blue stars, orange circles, and pink diamonds represent YSCs, eYSCs I, and eYSCs II, respectively. }
    \label{fig:massAgeScatter}
\end{figure*}

\section{Methodology}\label{sec:methods}

\subsection{The Two-Point Correlation Function}
We make use of the angular two-point correlation function (TPCF) $\tp$ to quantify the degree of YSC clustering at various scales in the galaxies, following the established methodology of \cite{grasha_15}, \cite{grasha17}, and \cite{menon21}.
 In this work we use the implementation in the \texttt{astroML} python package \citep{astroML}, which we describe below. 

The angular TPCF is defined as the probability above Poisson of finding two clusters with angular separation $\theta$ at some physical scale.
 This probability within a solid angle $\mathrm{d}\Omega$ is given as 
\begin{equation}
    \mathrm{d} P=N[1+\omega(\theta)] \mathrm{d} \Omega
\end{equation}
where $N$ is the average surface density of the cluster sample. 

For a randomly distributed sample, the TPCF is flat, i.e. $1 + \omega (\theta) = 1$, while for a sample showing clustering $1 + \omega (\theta) > 1$ at smaller scales
 with a general decreasing trend. A self-similar distribution is expected to produce a correlation function of the form 
\begin{equation}
    1 + \omega(\theta) = \left( \frac{r}{r_0}\right)^{\gamma}
\end{equation}
in which $r_0$ is the characteristic length scale on which clustering takes place and $\gamma$ describes the hierarchical ordering \citep[see][]{calzetti_89}. 

$\tp$ is calculated using the catalog of `real' data points (clusters) and a catalog of `false' data points generated randomly in position. 
The number of real pairs ($ \mathrm{DD(\theta)}$), false pairs ($\mathrm{RR(\theta)}$), and real-false pairs ($\mathrm{DR(\theta)}$) within some angular bin are computed, 
and $\omega(\theta)$ is estimated. \texttt{astroML} allows for the selection of one of several estimators for $\tp$. Here, we make use of the Landy-Szalay estimator 
\citep[][]{LS} which takes the form 
\begin{equation}
    \omega_{\mathrm{LS}}(\theta)=\frac{\mathrm{DD}(\theta)-2 \mathrm{DR}(\theta)+\mathrm{RR}(\theta)}{\operatorname{RR}(\theta)}.
\end{equation}
$ \mathrm{DD(\theta)}$, $\mathrm{DR(\theta)}$, and $\mathrm{RR(\theta)}$ are normalized to the total number of pairs in each catalog, calculated as 
\begin{equation}
    \mathrm{DD}(\theta)=\frac{N_{\mathrm{Pairs}, \ \mathrm{R}}}{N_{\mathrm{D}} N_{\mathrm{D}}},
\end{equation}
\begin{equation}
    \mathrm{DR}(\theta)=\frac{ N_{\mathrm{Pairs}, \ \mathrm{R+D}}}{N_{\mathrm{D}} N_{\mathrm{R}}},
\end{equation}
and
\begin{equation}
    \mathrm{RR}(\theta)=\frac{N_{\mathrm{Pairs}, \ \mathrm{D}}}{N_{\mathrm{R}} N_{\mathrm{R}}}
\end{equation}
in which $N_{\mathrm{Pairs}, \ \mathrm{R}}$, $N_{\mathrm{Pairs}, \ \mathrm{R+D}}$, and $N_{\mathrm{Pairs}, \ \mathrm{D}}$ are the number of false-false, 
real-false, and real-real pairs, respectively, and $N_{\rm D}$ and $N_{\rm R}$ are the total number of real and false points, respectively. 

For our analysis, we use 20 equally spaced angular separation bins, ranging from the resolution limits of the observations ($\sim\mathbf{5}$ pc), to the approximate size of the FOV. Angular separation bins which approach the size of the FOV are also susceptible to edge effects due to the diminishing number of pairs. \cite{menon21} showed that at $\sim\theta_{\rm max}/5$ and beyond, where $\theta_{\rm max}$ is the FoV size of observations, interpretations of results are susceptible to influence due to edge effects. We highlight this region when presenting the TPCFs (see, e.g., Figure \ref{fig:tpcfs}).

\subsection{Deprojection}
To avoid issues caused by the inclination of this sample galaxies, we deproject cluster positions from the plane of the sky to the plane of the galactic disk for NGC~4449 \citep[$i \approx 68^{\circ}$][]{hunter05}. Galaxies with inclination $i < 40^{\circ}$ experience minimal impact on the derived separations from inclination effects, hence we do not perform the deprojection on any of the other galaxies in this sample \citep{grasha_15}.

To perform the deprojection, the initial $x$ and $y$ positions in the plane of the galaxy are transformed such that 
\begin{equation}
    \begin{aligned}
& x^{\prime}=x \cos \theta+y \sin \theta \\
& y^{\prime}=\frac{-x \sin \theta+y \cos \theta}{\cos i}
\end{aligned}
\end{equation}
where the position angle $\theta$ is determined from the orientation of observed field of view, and $x^{\prime}$ and $y^{\prime}$ are the deprojected positions of the clusters.

\subsection{Model Fitting}\label{sec:33}
\begin{figure}
    \centering
    \includegraphics[width=\linewidth]{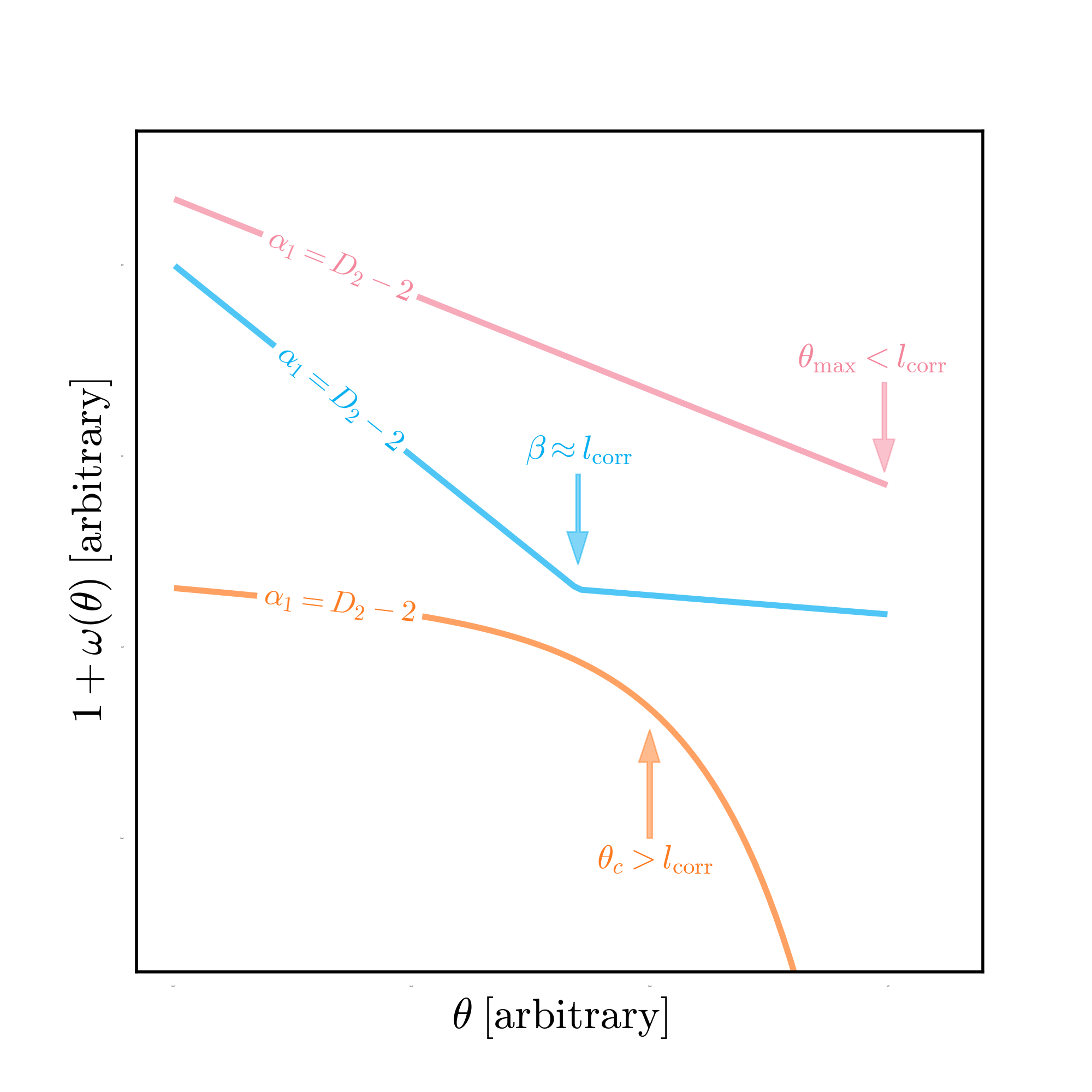}
    \caption{A brief overview of each model used in our analysis, along with quantities derived from each. The single, piecewise, and exponential truncation power models are shown in pink, blue, and orange, respectively. The fractal index $D_2$ is calculated from the inner slope of each of the models, while each model provides different constraints on the maximum scale of hierarchy $l_{\rm corr}$. Figure adapted from \cite{menon21}.}
    \label{fig:modelSummary}
\end{figure}

In order to quantify observed features in the TPCFs and derive physical quantities from them, we fit each computed TPCF to three models, selecting one best-fit model using a goodness-of-fit test. Our fitting routine is described in detail in Appendix \ref{sec:fitting}. The models used here are a single power law, a piecewise power law, and a power law with an exponential truncation. The models and fitting procedure are identical to those presented in \cite{menon21}. While previous work typically focused on only using single power law models to describe TPCFs, \cite{menon21} found that the three models described below can account for the underlying distributions implied by TPCFs, and provide a means to directly quantify  features and physical quantities inferred from TPCFs. Each of the models used in this analysis are visualized in Figure \ref{fig:modelSummary}, which also provides an overview of the physical quantities derived from each, described further in Section \ref{sec:pqs}. In the following, we describe each of the models. 

\subsubsection{Single Power Law Model}
The single power law model is given as 
\begin{equation}
    1 + \omega(\theta) = A \theta^{\alpha}
\end{equation}
where $A$ is the clustering amplitude, and $\alpha$ is the slope. The TPCF of a self-similar, 
fractal distribution is a single power law of the form $1 + \omega(\theta) \propto \theta^{\alpha}$ \citep{calzetti_89},
with the slope related to the two-dimensional fractal index such that $D_2 = \alpha + 2$. A population of star clusters which has a fractal and self similar underlying distribution can arise from the fractal nature of the natal gas clusters are born in \citep{elmegreen18}. The youngest cluster populations, which are closely linked to their natal gas, should align with this model as found by \cite{menon21}.

\subsubsection{Piecewise Power Law Model}
The second model, a piecewise power law, can be written as 
\begin{equation}
    1 + \omega(\theta) = \begin{cases}
        A_1 \theta^{\alpha_1} & \text{if}\;\; \theta < \beta,\\
        A_2 \theta^{\alpha_2}  & \text{if}\;\; \theta > \beta .
        \end{cases}
\end{equation}
in which $\beta$ is the transition point, and $\alpha_i$ and $A_i$ correspond to the amplitude and slope for
$\theta < \beta$ and $\theta > \beta$. The piecewise model represents cluster populations for which the underlying distribution is clearly self-similar up to some point $\beta$ where the underlying distribution becomes flatter if $\alpha_2 \to 0$. In practice, if $\alpha_2 \approx 0$, $\beta$ represents the maximum scale up to which the underlying distribution is fractal. Otherwise, the underlying distribution does not approach a truly random distribution, and instead remains non-uniform due to the underlying structure of the galaxy. 

\subsubsection{Power Law with Exponential Cutoff Model}
The third model we use for fitting is a power law with an exponential cutoff. Given as 
\begin{equation}
    1 + \omega(\theta) = A\theta^{\alpha} \exp\left(-\frac{\theta}{\theta_c}\right)
\end{equation}
with $\theta_c$ providing the angular scale at which the TPCF begins falling off exponentially. $A$ and $\alpha$ are defined the same as both other models, hence $D_2$ can be calculated as $\alpha + 2$. This model provides an upper limit on $l_{\mathrm{corr}}$ if  $l_{\mathrm{corr}} < \theta_{\rm c}$. This model tends to best represent the TPCFs of older cluster populations, reflecting the randomization of their distribution as a cluster population ages \citep{menon21}.

\section{Results \& Discussion} \label{sec:results}

\subsection{TPCFs}

\begin{figure*}
    \centering
    \includegraphics[width=\linewidth]{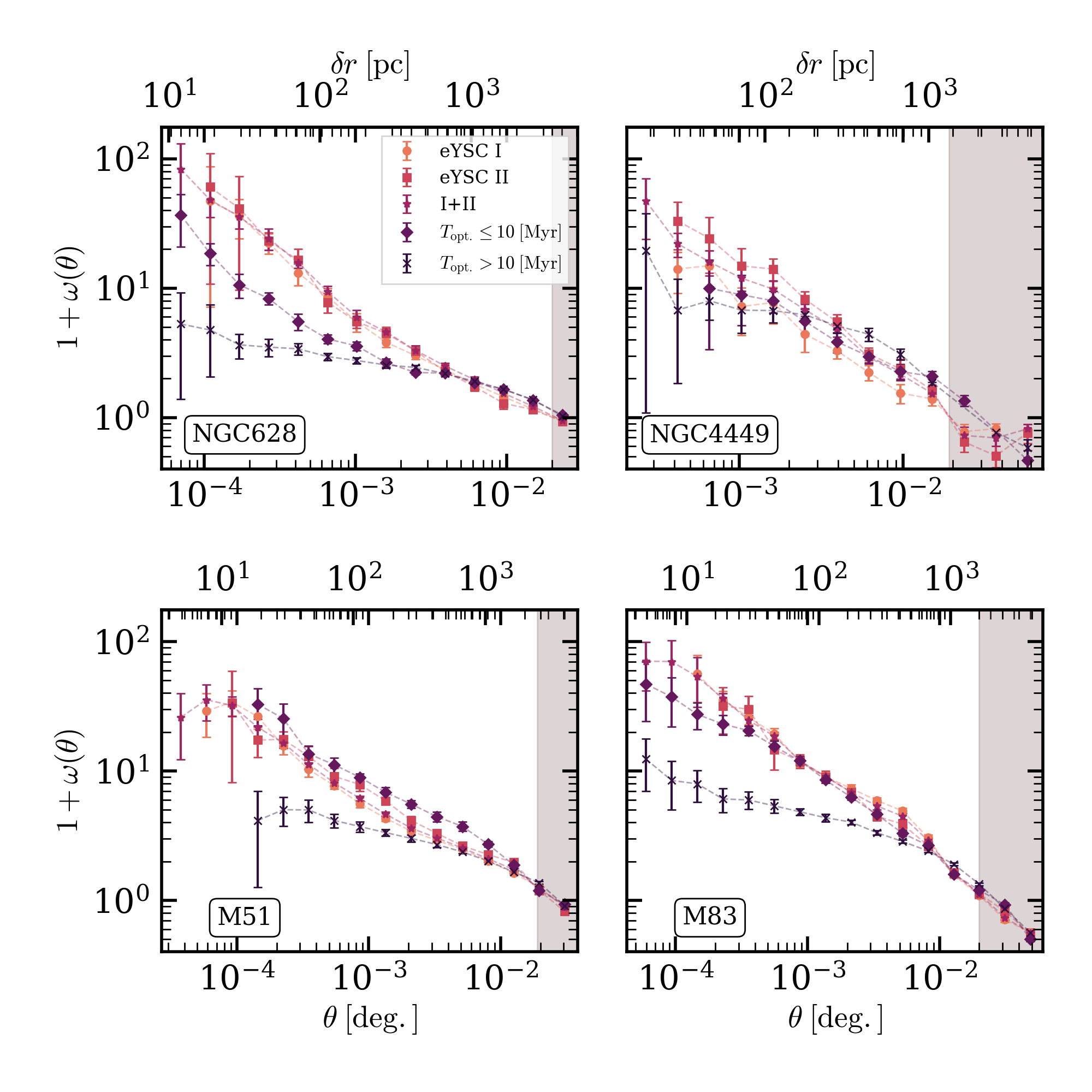}
    \caption{The TPCFs of five cluster samples in each of our four galaxies: eYSCs I, II, I+II (pink circles, blue squares, and orange stars, respectively), oYSCs with age $\leq$ 10 Myr (purple diamonds), and YSCs with age $>$ 10 Myr (magenta crosses). Points represent the TPCF calculated in that angular separation bin, and error bars are calculated via bootstrapping. The bottom axes provide angular separation scales, while those same scales are converted in physical units of pc for ease of interpretation. Shaded regions correspond to separation bins $> \theta_{\rm max}/5$ where edge effects may influence results (see discussion in \cite{menon21}).}
    \label{fig:tpcfs}
\end{figure*}

Figure \ref{fig:tpcfs} presents the calculated TPCFs from the spatial distribution of eYSC Is, eYSC IIs, combined eYSC classes, oYSCs with age $\leq 10$ Myr, and YSCs with age $> 10$ Myr for each of the four galaxies in this sample. Reported error bars are calculated via bootstrapping. We ensure this approach provides adequate measure of error in Appendix \ref{sec:errorTest}. TPCFs are calculated across 20 angular separation bins, corresponding to separations of 1-5000 pc, spaced logarithmically. We mask out angular separation bins for which $1+\omega(\theta) \leq 0$, and in bins where $\omega(\theta)$ is smaller than the bootstrap calculated error, as both of these situations arise due to a low number of pairs in a given angular separation bin. We also provide a physical scale in units of parsecs for ease of interpretation. 

Several trends are immediately apparent from Figure \ref{fig:tpcfs}. First, the TPCFs of eYSC populations exhibit near identical degrees of clustering in each of our galaxies. As the separation in age between eYSC Is and IIs is expected to be small, this is not surprising. Now looking at the three spirals in this sample (NGC 628, M51, M83), the $>10$ Myr YSC populations are all nearly flat, indicating little-to-no spatial clustering. This implies a randomization timescale of order $\sim 10$ Myr, aligning with previous findings \citep{grasha_15, grasha17, grasha18, grasha19, menon21, turner22}. TPCFs of the young oYSC population ($\leq 10$ Myr) in M51 and M83 are very similar to the TPCFs of the eYSC populations, while in NGC 628 there is a clear difference between the TPCFs. This finding implies that NGC 628 has a somewhat shorter randomization timescale than the other two spirals. 

NGC 4449, the only dwarf in this sample, appears to have TPCFs with very similar clustering amplitudes across the populations. There is however a hint that the perfect self-similarity breaks down at the oldest ages: while the eYSCs and young/intermediate YSCs (out to 100 Myr) align well, the oldest YSC TPCF, those older than 100 Myr, have a shallower slope (see Table \ref{tab:models}). This may be due to the lack of shear in NGC 4449 compared to the spirals, leading to less dramatic spatial randomization across YSC  populations.

Qualitatively, the TPCFs all seem to imply that the spatial distribution of clusters becomes randomized on timescales of $\gtrsim 10$ Myr. The eYSCs and young oYSCs share very similar spatial distributions but are both noticeably different than the old YSC population. This aligns with previous results, all of which have found significant variation between the TPCFs of young and old clusters. Older clusters in all of the galaxies have TPCFs that are generally flat at smaller separations and taper off at larger separations. As clusters populations age, they disperse throughout their host galaxy, forgetting the hierarchical structure imparted on the population at birth. In the following sections, we further expand upon the temporal evolution of the TPCFs, and quantify the differences between the different cluster populations across this sample.

\subsection{Temporal Evolution of TPCFs}

\begin{figure*}
    \centering
    \includegraphics[width=\linewidth]{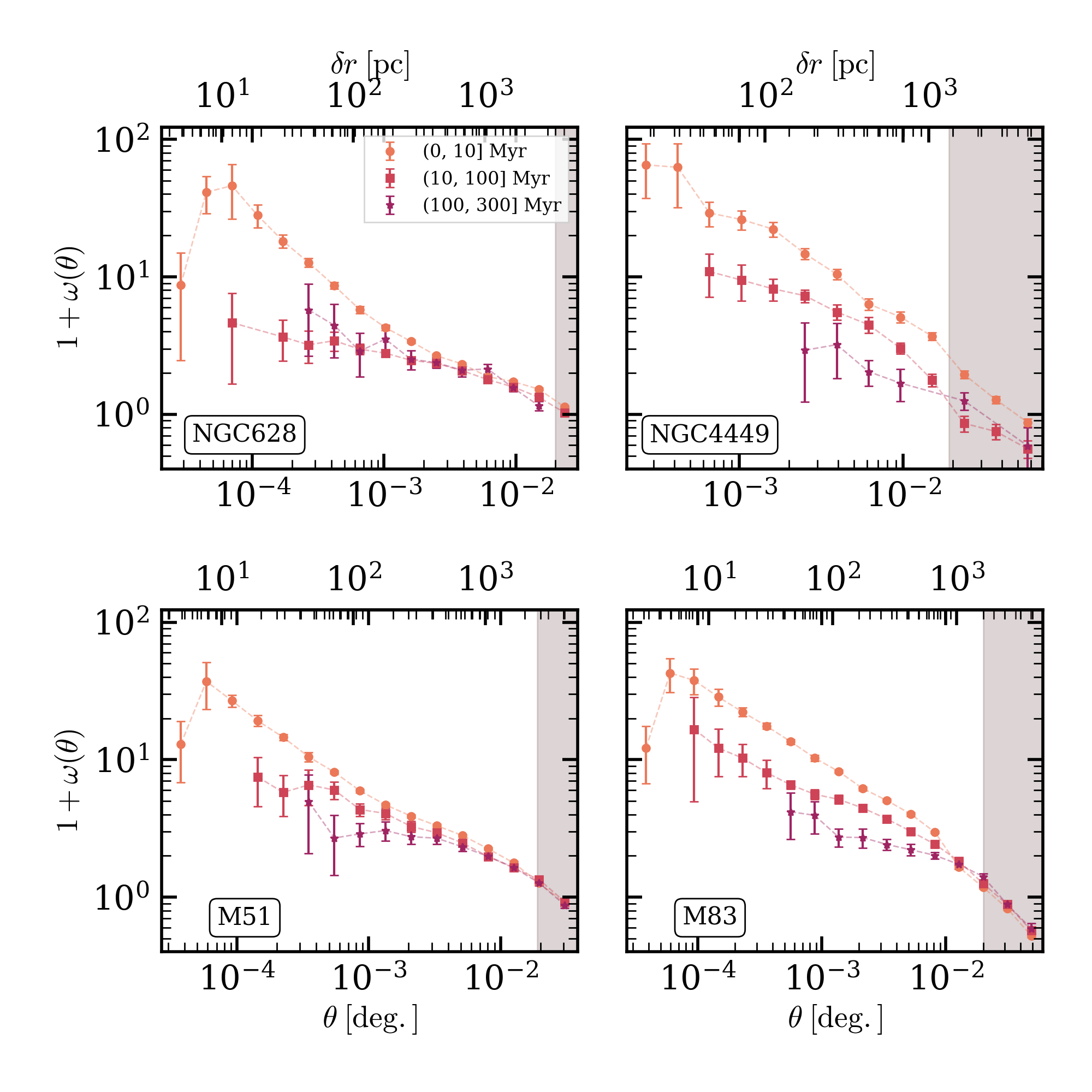}
    \caption{TPCFs of the complete cluster catalogs (eYSCs + YSCs) binned by age into three bins corresponding to ages in the ranges (0, 10] Myr (pink circles), (10, 100] Myr (blue squares), and (100, 300] Myr (orange stars). Points are calculated values, and error bars are calculated via bootstrapping. Bottom and top $x$-axes are the same as in Figure \ref{fig:tpcfs}. }
    \label{fig:ageBinTPCF}
\end{figure*}

To further investigate changes due to the time evolution of a given cluster population, we make three age cuts to our cluster catalogs. We bin a combined catalog of eYSCs and YSCs into $(0, 10],\;(10, 100],\;\mathrm{and}\;(100,300]$ Myr age bins, then compute the TPCFs of each age bin across our galaxy sample. None of the eYSCs have ages $> 10$ Myr, hence the final two age bins are comprised of only YSCs. We use the same bins and masking routine described above. These TPCFs are presented in Figure \ref{fig:ageBinTPCF}. 

Across this sample, the youngest age bins have TPCFs that show high levels of spatial correlation. The middle age bins all exhibit a significantly smaller level of spatial clustering, but are generally more clustered than the oldest age bin. The exception to this is NGC 628, where the middle and oldest age bins exhibit very similar clustering amplitudes, implying that spatial randomization occurs at $ \lesssim$ 10 Myr. This change in randomization timescale compared to the other spirals in our sample may be due to the fact that NGC 628 has weaker disk dynamics compared to M51 and M83, which have strong-arm disks and a central bar, respectively. While the eYSC and YSC TPCFs of NGC 4449 exhibit very similar behavior in Figure \ref{fig:tpcfs}, the age binned samples have distinct behavior, aligning with the spirals in this sample. 

The behavior of the TPCFs in Figure \ref{fig:tpcfs} implies that, as clusters age, their spatial distribution approaches a random distribution. This is shown clearly and hence confirmed by the age binned, all cluster results in Figure \ref{fig:ageBinTPCF}. Previous work has shown very similar results \citep{odekon06, sanchez09, grasha_15, grasha17, grasha18, grasha19, menon21, turner22}.

There is a noticeable dip in the TPCFs present in the youngest age bin for each of the spiral galaxies in this sample, occuring at separation scales $<$ 10 pc. %At small separation scales, the 
In theory, the TPCFs of cluster populations is expected to produce this feature due to the impact of random motions of the clusters themselves, and should only appear in cluster populations with a fine temporal separation in the formation \citep[see][]{elmegreen18}. However, 
%by design, 
because of angular resolution limitations, our YSC catalogs are not complete for the separations at which this feature is seen. While theoretically expected, this feature is likely the result of incompleteness at small spatial separations, and not a true detection.

\subsection{Quantifying the Behavior of TPCFs}

\begin{figure*}
    \centering
    \includegraphics[width=\linewidth]{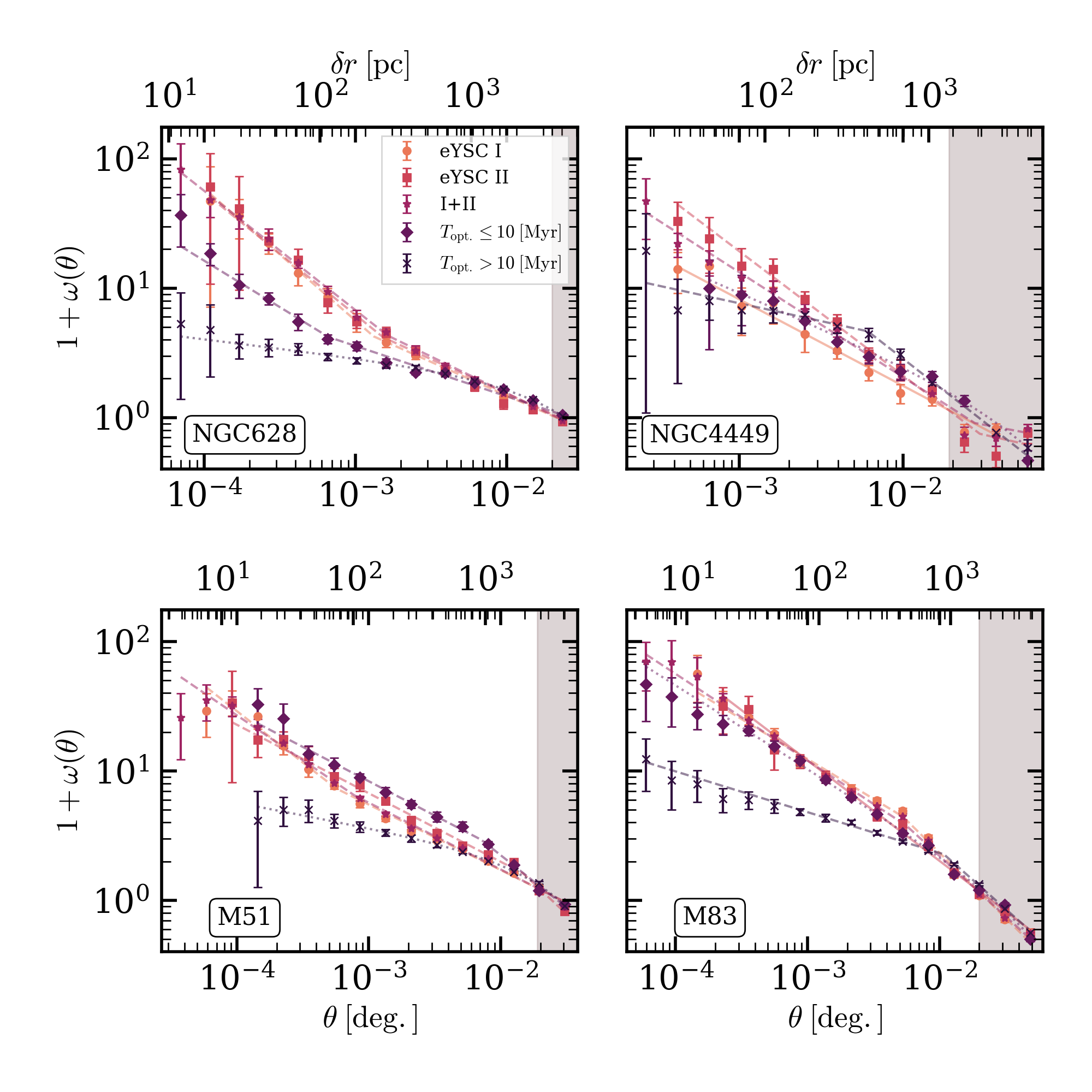}
    \caption{The same as Figure \ref{fig:tpcfs}, but with the best fit model over plotted alongside its corresponding TPCF in the same color. Solid lines correspond to the single power law model, dashed lands the piecewise power law model, and dotted lines to the exponential truncation model. }
    \label{fig:tpcfsBestFit}
\end{figure*}

\begin{figure*}
    \centering
    \includegraphics[width=\linewidth]{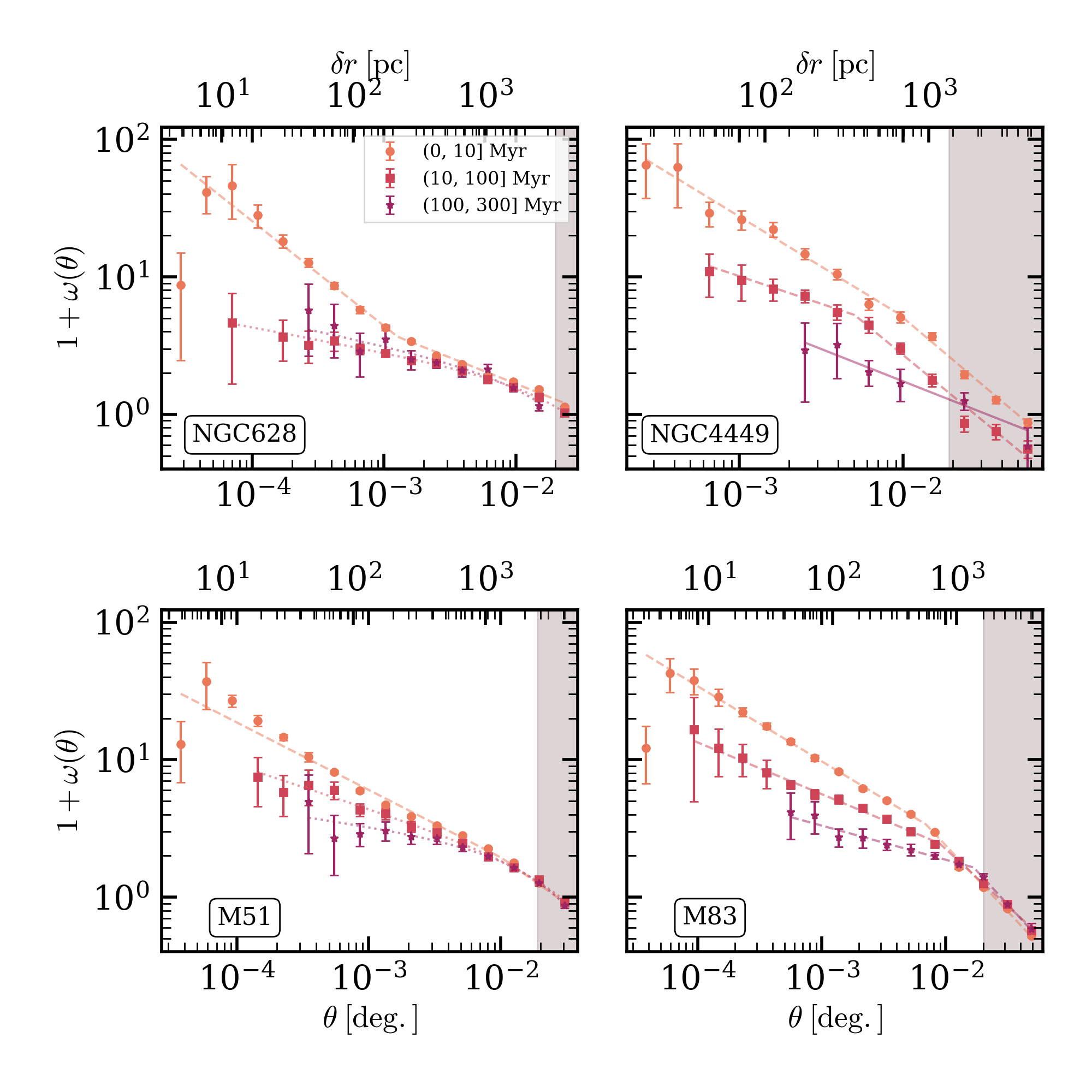}
    \caption{The same as Figure \ref{fig:ageBinTPCF}, but with the best fit model over plotted alongside its corresponding TPCF in the same color. Solid lines correspond to the single power law model, dashed lands the piecewise power law model, and dotted lines to the exponential truncation model.}
    \label{fig:tpcfsBestFitAgeBin}
\end{figure*}

To further solidify the results discussed in the previous two sections, we also present best-fit models to each of the TPCFs shown in Figures \ref{fig:tpcfs} and \ref{fig:ageBinTPCF} in Table \ref{tab:models}. We also visualize best-fit models alongside the corresponding TPCFs in Figures \ref{fig:tpcfsBestFit} and \ref{fig:tpcfsBestFitAgeBin} for the eYSC/oYSC and all cluster age binned TPCFs, respectively. 

In a given galaxy, the TPCFs of different eYSC samples all tend to have similar best-fit values of the inner power law slope $\alpha_1$. Nearly all of the eYSC populations are best fit by the piecewise power law model. The exceptions are the eYSC I bin in NGC 4449 and the eYSC II bin in M83, which are both fit by single power laws. The eYSC models in NGCs 628 and 4449 have very prominent breaks in the power law, with the break in NGC 4449 happening near the edge of the angular separation bins. This could possibly be due to the influence of edge effects. The power law breaks are much less prominent in M83 and M51, implying a moderate deviation from the underlying fractal distribution at larger scales.

YSCs with ages $ > 10$ Myr are best fit by the exponential truncation model in NGC 628 and M51. In NGC 4449 and M83, this bin is best fit by a broken power law, with $\alpha_2 < \alpha_1$. The exponential truncation model implies that the oldest star clusters approach a radially distributed random distribution as they age \citep[see discussion in ][]{menon21} in the spiral galaxies. In the case of NGC 4449 and M83, it may be the case that edge effects limit our ability to detect the truncation due to those galaxies being nearest, and hence having the smallest FoVs. In NGC 4449, the old age bin YSCs still show some degree of fractality in their underlying distribution, possibly due to the irregular morphology of the galaxy. The lack of shear in NGC 4449 possibly maintains this fracticality, consistent with the similar amplitudes shown in Figure \ref{fig:tpcfs}. Indeed, \cite{renaud24} find that shear decreases with decreasing gas fraction, consistent with this picture for a metal poor starburst galaxy like NGC 4449.

Looking at the combined eYSC+YSC age bins, the youngest age bin (0, 10] Myr TPCFs are all best fit by piecewise power laws. NGC 4449, and M83 again have $\alpha_2 < \alpha_1$ in this bin, implying that these populations may be representative of a truly hierarchical distribution up to larger scales, and that the combination of edge effects and limited FoV is again influencing our results.

The $(100, 300]$ Myr bin is best fit by a variety of models across the sample. In NGC 628 and M51, this age bin is best fit by the truncated model, while NGC 4449 and M83 are best fit by single and piecewise power laws, respectively. This implies the oldest cluster populations in these galaxies still exhibit some degree of an underlying hierarchical ordering, however both have smaller amplitudes compared to their $(10, 100]$ Myr age bin TPCFs. Interestingly, only M51 and NGC 628 imply a nearly true Poisson distribution in this age bin, along with the $(0, 100]$ Myr bin in NGC 628. In M51, the relatively strong shear and/or the interaction with its companion galaxy may be driving the randomization of the spatial distribution. M83 and NGC 4449 both show clear signs of non-uniformity in the spatial distributions of the $(100, 300]$ Myr bin. \cite{kim24} find that bar driven shear can inhibit star formation in massive galaxies such as M83, which may explain the longer randomization timescales implied by the TPCFs. \cite{ni25} also find that in simulations, shear plays an important role in regulating star formation at cloud scales, aligning with our findings here.

\input{Table2}

\begin{figure}
    \centering
    \includegraphics[width=\linewidth]{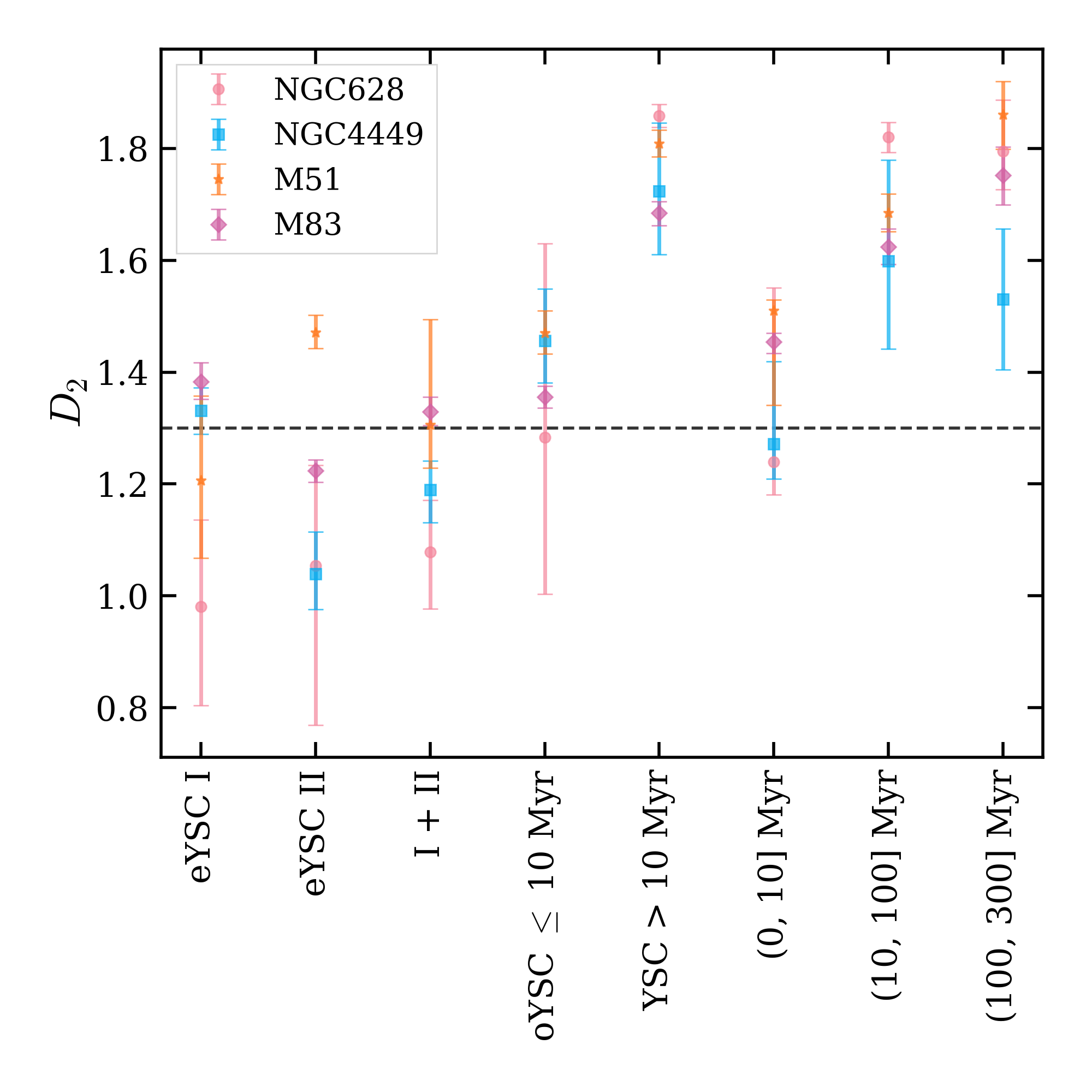}
    \caption{The distribution of $D_2$ values obtained from fits to the TPCFs for each population of clusters across this sample of galaxies. Cluster populations are labeled on the x-axis. NGC 628, NGC 4449, M51, and M83 correspond to orange circles, red squares, purple stars, and indigo diamonds, respectively. Error bars are the 14th and 84th percentile errors. The dashed line corresponds to $D_2 = 1.3$, the expected value of a turbulent-driven ISM setting the underlying distributions of clusters.}
    \label{fig:D2Dist}
\end{figure}

\subsection{Physical Quantities from TPCFs} \label{sec:pqs}
The final portion of our analysis focuses on obtaining physical quantities $D_2$ and $l_{\rm corr}$ from the TPCFs. Parameter $D_2$ represents the fullness of the underlying distribution. A value of $D_2 = 2$ represents a completely random distribution (points evenly distributed across a plane). Correlation length, $l_{\rm corr}$, describes the scale where the nature of the underlying distribution shifts from hierarchical to random or has a change in the degree of underlying hierarchy. $l_{\rm corr}$ should align with the largest gas structures in a given galaxy \citep{efremov95} (i.e., the largest scale up to which the spatial distributions are tracing the underlying fractal ISM), while $D_2$ has been proposed to hold a universal value of $D_2 \sim 1.3$ if the star formation process is universal \citep[e.g.,][]{feitzinger87, elmegreen_falgaron_96}. That is, if the self-similar hierarchy observed is set self-consistently by turbulence in the ISM, $D_2$ should be universal \citep{elmegreenscalo04}. In Table \ref{tab:physicalquants}, we provide both of these quantities, along with their 16th and 84th percentile errors, for each of the 8 cluster populations across our galaxy sample. These results are also visualized in Figures \ref{fig:D2Dist} and \ref{fig:lcorrDist}.

\subsubsection{Two-dimensional Fractal Indices}
Generally, values of $D_2$ derived from the TPCFs here increase with older cluster populations across this sample of galaxies. eYSCs and oYSCs with ages $\leq$ 10 Myr tend to have similar values of $D_2$, while the $>$ 10 Myr YSC populations have noticeably higher values, except in NGC 4449. The eYSC + YSC age binned populations all have agreeable results across the populations in each of our galaxies, and exhibit a clear increase between the youngest age bin and the middle/oldest age bins. This further aligns with the framework of hierarchical cluster populations dispersing as the population ages. 

Recent studies that derive $D_2$ from the spatial distributions of young star forming clumps \citep{shashank25} and YSCs detected in the UV/optical \citep{shashank25, menon21, grasha17} find significant variation in $D_2$ across their samples of galaxies, challenging the proposed theoretical value, and implying some dependence on the star formation process to galaxy-wide properties. It has generally been found that $D_2 \in [0.9, 1.5]$ for local-volume star forming galaxies \citep[e.g.,][]{elmegreen06, scheepmaker09, grasha17, menon21, shashank25}.

Young clusters are more closely coupled to the natal gas they arise from, and should act as more robust tracers of the self-similar and hierarchical distribution they are born from. The youngest populations in this sample, the eYSCs, young oYSCs, and all clusters with ages $\leq$ 10 Myr, all produce values of $D_2$ which align remarkably well with each other, generally agreeing within a $\sim 1.5\sigma$ error. This contrasts with the findings of \cite{menon21} and \cite{shashank25}, who both find significant variation in $D_2$ from galaxy to galaxy. this sample, although only consisting of four galaxies, has significantly better number statistics than \cite{menon21} and \cite{shashank25}, and includes eYSCs that are still embedded in their natal clouds, and subsequently more intimately linked to the underlying fractal ISM. This should, in theory, result in better estimations of $D_2$. Due to this and having the best number statistics, the $(0, 10]$ Myr age bin should provide the best measurement of $D_2$. In this bin, we find for NGC 628, NGC 4449, M51, and M83 that $D_2 \approx \{1.24, 1.27, 1.51, 1.45\}$, respectively (Fig \ref{fig:D2Dist}).

Recently, NGC 628 has been found to have much lower values of $D_2$ compared to similar galaxies (e.g., \cite{menon21} found that $D_2 \approx 0.9$). We find that $D_2 = 1.24^{+0.312}_{-0.058}$, agreeing with the theoretical expectation, but also tending towards a higher value. This suggests that previous measurements may have been biased due to incompleteness in the sample, and the fact that oYSCs are inherently less coupled to the ISM compared to our eYSC catalogs. The two other spirals, M51 and M83, have slightly higher values of $D_2$, with $D_2 = 1.51^{+0.0197}_{-0.168}$ and $1.45^{+0.0157}_{-0.0203}$, respectively. While higher than the expected $D_2 \approx 1.3$, these values are notably not substantially higher or lower than $\approx 1.3$. Slight deviations from $\approx 1.3$ may be due to a variety of galaxy specific factors (e.g., M51 is actively interacting with a companion, and M83 has a central bar). We find that $D_2 =1.27^{+0.147}_{-0.0631}$ in NGC 4449, completely consistent with the theoretically expected value.

\subsubsection{Maximum Scale of Hierarchy}
We find a range of $l_{\rm corr}$ across this sample. The youngest populations of clusters should have underlying distributions which most closely reflect that of the underlying structure of the ISM, hence we take the youngest YSC age bin to provide the best values of $l_{\rm corr}$. Values for other cluster populations are provided for sake of completeness. 

\begin{figure}
    \centering
    \includegraphics[width=\linewidth]{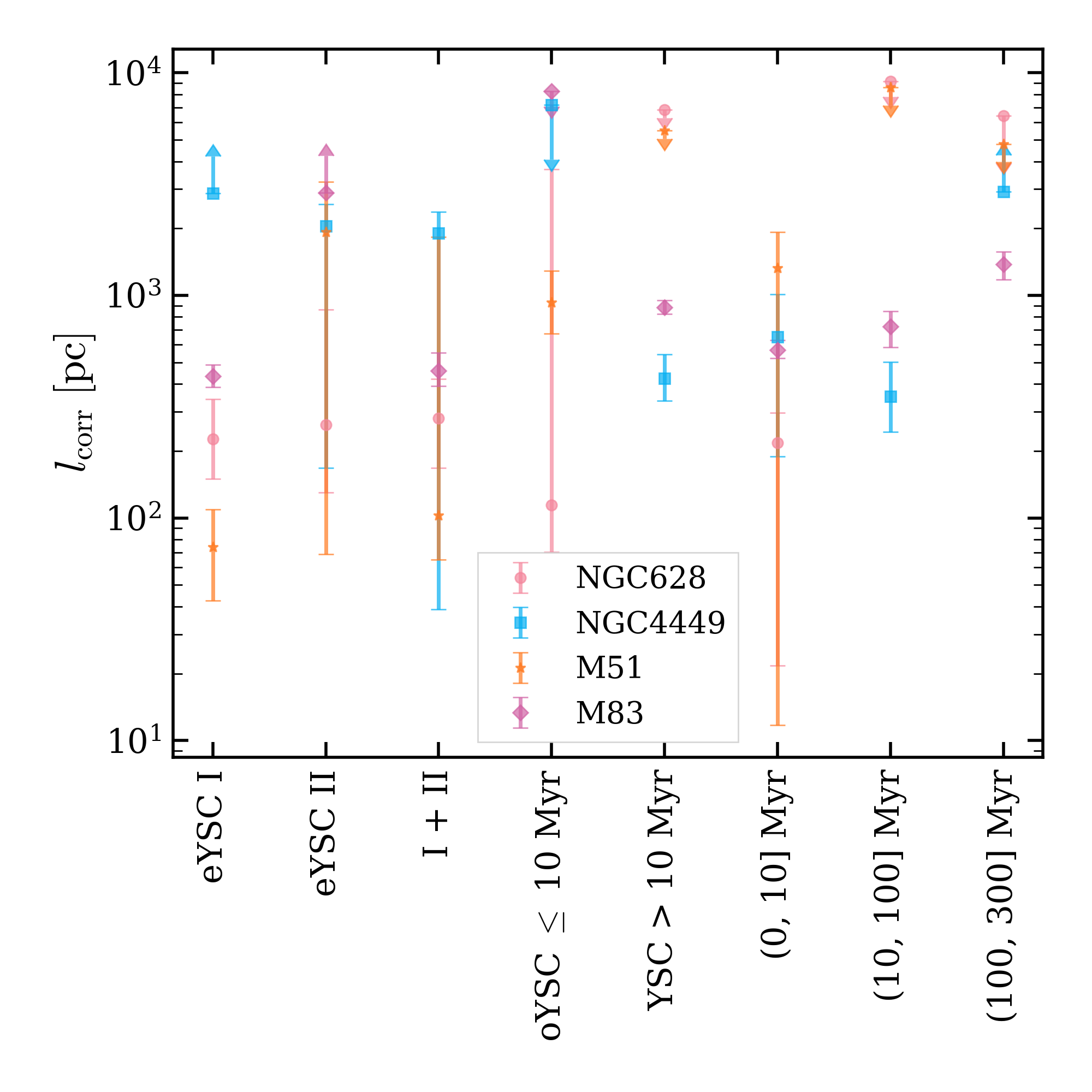}
    \caption{The distribution of $\lcorr$ values derived from TPCFs for each cluster population across this sample of galaxies. Upwards and downwards arrows indicate lower and upper limits, respectively. Different cluster populations are labeled on the x-axis. NGC 628, NGC 4449, M51, and M83 correspond to orange circles, red squares, purple stars, and indigo diamonds, respectively. Error bars are the 14th and 84th percentile errors.}
    \label{fig:lcorrDist}
\end{figure}

In the spiral galaxies, we find vales of $l_{\rm corr}$ which range from a few hundred to $\sim$ 1000 pc. In most cases when $\lcorr$ is estimated from the piecewise power law model, the transition in slopes at $\beta$ is minimal at best (M51) or the model has steeper $\alpha_2$ in the regime where edge effects may influence the estimated values of the TPCF (M83 and NGC 4449). NGC 628 is the only galaxy in which the youngest age bin exhibits a clear visual break in the TPCF, hence provides what can be considered as the only good estimate of $\lcorr$. This is potentially due to it being the most distant galaxy in this sample, hence edge effects are less prominent. The youngest age bin for all clusters agrees very well with values produced by the eYSC populations, with $\lcorr \approx 210-300$ pc in these cases. This aligns well with a recent value of $\lcorr = 190^{+70}_{-40}$ reported by \cite{menon21}. Values estimated from older cluster populations generally provide constraints which agree among themselves, with each of the galaxies exhibiting clear differences in values with the others (see Figure \ref{fig:lcorrDist}).  

The lack of clear visual breaks in the other galaxies may be due to edge effects. That is, the distributions are likely to be representative of a truly hierarchical, scale-free distribution (which should produce single power law TPCFs), but the influence of edge effects results in TPCFs which are best fit by piecewise power law models. This suggests that in NGC 4449, M51, and M83, the distributions remain hierarchical beyond the derived values of $l_{\rm corr}$.

\input{Table3}

\section{Summary \& Conclusions} \label{sec:conclusions}

In this work, we build on previous results and  established methodology by applying the angular two-point correlation function to the spatial distribution of YSCs in several nearby galaxies. We extend previous work by using new eYSC and YSC catalogs produced as part of the JWST-FEAST project. By including emerging-phase IR-detected YSCs and increasing number statistics, we trace the ISM’s hierarchical structure at younger timescales, which are more directly linked to cluster formation, than was possible with HST-only catalogs.

Across a sample of three spiral galaxies (NGC 628, M51, M83) and one dwarf (NGC 4449), we separate the star cluster catalogs into several bins according to SED-derived ages, or morphological classification (see Sections \ref{sec:eyscclass}, \ref{sec:oyscClass}). We compute TPCFs for each bin and determine which of three models best fits the TPCF using a MCMC-oriented approach. Our main findings are summarized as:

\begin{itemize}
    \item oYSCs with ages $\leq 10$ Myr and eYSCs have TPCFs which imply very similar underlying fractal distributions across this sample of galaxies. eYSCs I, II and a combined population of both have TPCFs which have very little difference. YSCs with ages $>$ 10 Myr have TPCFs which exhibit little-to-no spatial correlation, implying that clusters populations detected in the optical begin to become randomly distributed at ages $\gtrsim$ 10 Myr. 
    \item Generally, we find that TPCFs of clusters with ages in the range (0, 10] Myr exhibit clear behavior of having underlying hierarchical distributions in space, in agreement with previous results from \cite{grasha_15, grasha17, grasha19, menon21}. Older clusters with ages in the range (100, 300] Myr have TPCFs which show a clear deviation from this behavior, and imply nearly Poisson underlying distributions, interpreted as the randomization of the distribution as the clusters disperse throughout their host galaxy. Clusters with ages in the range (10, 100] tend to still exhibit some degree of hierarchical structure in their underlying distributions, but noticeably less so than younger cluster populations.
    \item As an exception to the above, older cluster populations in NGC 4449 still exhibit a clear hierarchical ordering. We propose that this may be due to the lack of shear experienced by the galaxy, as it is the only dwarf in this sample.
    \item Derived values of the two-dimensional fractal index $D_2$ tend to agree for the youngest and oldest cluster populations. The youngest populations produce values of $D_2$ which align better with the prediction of $D_2 \sim 1.3$ for a universal star formation process than those reported in previous works. Clusters with ages in the range (10, 300) Myr have values of $D_2$ which approach the value of a Poisson (random) distribution. 
\end{itemize}

Our analysis seems to hint towards a uniform picture of hierarchical star formation across galaxies. Future work leveraging other analysis methods for spatial point processes (e.g., an inhomogeneous $K$-function approach) could expand on results presented here. Cluster populations across all of our spirals indicate a moderate level of similarity in their underlying hierarchical structure. The only dwarf in this sample, NGC 4449, is a clear outlier in the behavior of its TPCFs. In particular, hierarchical clustering in this galaxy appears to persist beyond $\sim 100$ Myr, much older ages than for the spirals; this behavior may possibly be due to the much weaker shear present in the dwarf, which will enable preservation of the clustering for longer times than in the spirals.  Its different morphology, and worse number statistics, makes direct comparisons between this system and the other galaxies challenging. Future work using eYSC catalogs in more dwarf galaxies may reveal similar trends in that population of galaxies, and continue pointing towards a somewhat homogeneous picture regarding the hierarchical nature of star formation.

%% Please use the acknowledgment and contribution environments. This will 
%% be anonomyized when the "anonymous" style option is used. 
\begin{acknowledgments}
The authors thank the anonymous reviewer for insightful feedback which greatly improved the quality of this manuscript. D.L. and D.C. thank the National Science Foundation for supporting this work through  %D.L. was supported by the National Science Foundation under
Grant \# 2406687.  D.L. thanks Sophia Kalakailo, Joseph Golec, Massissilia Hamadouche, Benjamin N. Velguth, Aidan Cloonan, and Sinclaire Manning for helpful discussions. K.G. is supported by the Australian Research Council through the Discovery Early Career Researcher Award (DECRA) Fellowship (project number DE220100766) funded by the Australian Government. A.A acknowledges support from Vetenskapsr\aa det
2021-05559. A.A and A.P. acknowledge support from the Swedish National Space Agency (SNSA) through the grant 2021- 00108. A.A. and H.F.V. acknowledges support from (SNSA) 2023-00260. E.S. is supported by the international Gemini Observatory, a program of NSF NOIRLab, which is managed by the Association of Universities for Research in Astronomy (AURA) under a cooperative agreement with the U.S. National Science Foundation, on behalf of the Gemini partnership of Argentina, Brazil, Canada, Chile, the Republic of Korea, and the United States of America. ASMB acknowledges the support from the Royal Society University Research Fellowship URF/R1/191609. GÖ acknowledges support from the to Swedish National Space Agency (SNSA) and the Swedish Research Council (VR). MM acknowledges financial support through grants PRIN-MIUR 2020SKSTHZ, the INAF GO Grant 2022 “The revolution is around the corner: JWST will probe globular cluster precursors and Population III stellar clusters at cosmic dawn,” and by the European Union – NextGenerationEU within PRIN 2022 project n.20229YBSAN - "Globular clusters in cosmological simulations and lensed fields: from their birth to the present epoch”
\end{acknowledgments}

\facilities{}

\software{Astropy \cite{astropy}, astroML \cite{astroML}, Matplotlib \cite{Hunter:2007}, Numpy \cite{harris2020array}, SciPy \cite{2020SciPy-NMeth}, emcee \cite{emcee}}


\appendix

\section{Impact of Completeness on Calculated TPCFs}\label{sec:completeness}

\begin{figure*}
    \centering
    \includegraphics[width=\linewidth]{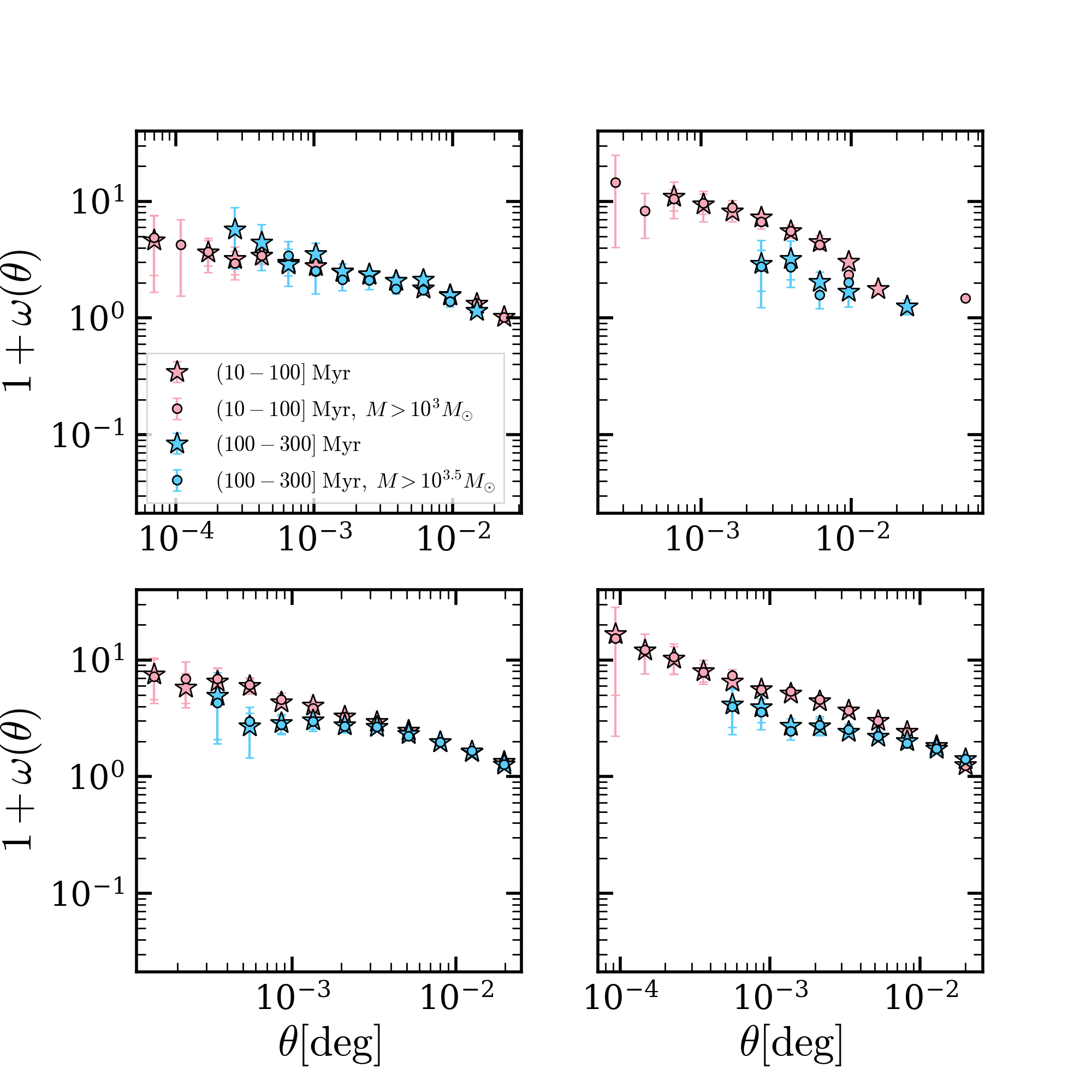}
    \caption{The TPCFs of YSCs in the (10-100] (pink) and (100-300] (blue) Myr age bins. Stars are the same TPCFs shown in Figure \ref{fig:ageBinTPCF}, while circles are TPCFs computed after cutting all clusters with masses $> 10^3$ M$_{\odot}$ in the (10-100] Myr age bin (pink circles), and masses $> 10^{3.5}$ M$_{\odot}$ in the (100-300] Myr age bin (blue circles). }
    \label{fig:massTest}
\end{figure*}

The FEAST eYSC catalogs are complete down to $\sim 1000$ M$_{\odot}$, however older age bins, specifically the $(10-100]$ and $(100-300]$ Myr bins suffer from incompleteness below $\sim 10^{3.5}$ $M_{\odot}$, respectively. To confirm that completeness in these bins does not drastically affect our TPCFs, we re-compute the TPCFs for the $(10-100]$ and $(100-300]$ Myr age bins using YSCs with masses $> 10^3$ and $>10^{3.5}$ $M_{\odot}$ respectively. We show the mass cut TPCFs along with those of the full sample in Figure \ref{fig:massTest}. The mass cut TPCFs are nearly identical to those using the original TPCFs, hence the incompleteness of the older age bins has no major impacts on our results and analysis. 

\section{Fitting Routine}
\label{sec:fitting}

To fit each of the models described in Section \ref{sec:33} and determine which provides the best fit to a given TPCF, we employ a Markov-Chain Monte Carlo (MCMC) approach. For the single power law model, we fit for the parameter vector 
$\lambda = (A, \alpha)$. In the case of the broken power law, we fit for  $\lambda = (A_1, \alpha_1, \alpha_2)$ under the assumption that the amplitude is continuous, i.e., $A_1 = A_2 \beta^{\alpha_1 - \alpha_2}$. The exponential cutoff model is fit with $\lambda = (A_1, \alpha_1, \theta_c)$.

We fit using a likelihood function of the form 
\begin{equation}
    \mathcal{L} = - \frac{1}{2}\sum \left(\frac{D(\theta) - M(\theta | \lambda)}{\sigma_D(\theta)}\right)^2
\end{equation}
in which $D(\theta)$ is simply $1+\omega(\theta)$ for a given bin $\theta$, $M(\theta|\lambda)$ is the 
model value at that separation bin, and $\sigma_D(\theta)$ is the error in $1+\omega(\theta)$ for 
separation bin $\theta$. The priors for the models are as follows: $-10 <A_1<10$, $-5 < \{\alpha_1, \alpha_2\} \leq 0$, $\theta_{\rm min.} < \beta < \theta_{\rm max}$, and $\theta_{\rm min.} < \theta_{\rm cut} < \theta_{\rm max.}$. Here, $\theta_{\rm min}$ and $\theta_{\rm max}$ are the smallest and largest angular seperation bins used for fitting, respectively. For the fitting,
we make use of \texttt{emcee} \citep{emcee}, a python-based MCMC code. We run the MCMC for a total 5200 steps with 500 walkers, discarding the first 200 as burn-in. Convergence of parameters is visually confirmed.

To determine which model best fits a given TPCF, we compute a modified Akaike Information Criterion (AIC) \citep{akaike} for each models best-fit to a TPCF. Here we use the corrected AIC from \cite{CLIFFORDM.HURVICH1989Rats}, which takes the form
\begin{equation} \label{eq:aic}
    \mathrm{AIC} = 2N_{\lambda} - 2 \ln(\mathcal{L}_{\rm max})+\frac{2N_{\lambda}(N_{\lambda}+1)}{N_{\theta - N_{\lambda}-1}}
\end{equation}
in which $N_{\lambda}$ is the number of parameters being fit, $\mathcal{L}_{\mathrm{max.}}$
is the maximum of the likelihood function, and $N_{\theta}$ is the number of bins that $1 + \omega(\theta)$
is calculated over. The model which minimizes Eq. \ref{eq:aic} is deemed the best fit model for a given TPCF. 

\section{Testing Error Estimation of TPCFs}\label{sec:errorTest}

We perform additional tests to ensure that our estimation of error (via bootstrapping) is viable for the fitting of deterministic functions to the TPCFs. To do this, we compute the TPCF across 100 iterations of eYSC I positions, with each iteration being subject to a random perturbation in its RA and Dec. The distribution of resulting TPCFs is then compared to the values presented above.

Spatial perturbations are computed assuming a maximum shift in $(x, y)$ of 20 pcs. A maximum of 20 pc is chosen under the assumption that, if a star cluster has some random bulk motion $\sim 2$ km s$^{-1}$ with maximum age $\sim 10$ Myr, then it will move a maximum of $~20$ pc from its birth site. That is, we shift each pair of RA and Dec by some random values $\delta x$ and $\delta y$, where $20\;\mathrm{pc} \geq \sqrt{\delta x^2 + \delta y^2}$. For some random angle $\theta \in [0, 2\pi)$ and total shift $s = 20\;\mathrm{pc}  \times a$ with $a \in [0, 1)$, then $\delta x$ and $\delta y$ are calculated as 
\begin{equation}
    \begin{aligned}
& \delta x = s \cos(\theta)  \\
& \delta y = s \sin(\theta)
\end{aligned}
.
\end{equation}
For each of the 100 iterations, the $i$th point is shifted such that 
\begin{equation}
    \begin{aligned}
& x^{'}_i = x_i+\delta x_i = x_i + s_i \cos(\theta_i)   \\
& y^{'}_i = y_i+\delta y_i = y_i + s_i \sin(\theta_i) 
\end{aligned}
.
\end{equation}
for random $\theta_i$ and $s_i$ as described above.

\begin{figure*}
    \centering
    \includegraphics[width=\linewidth]{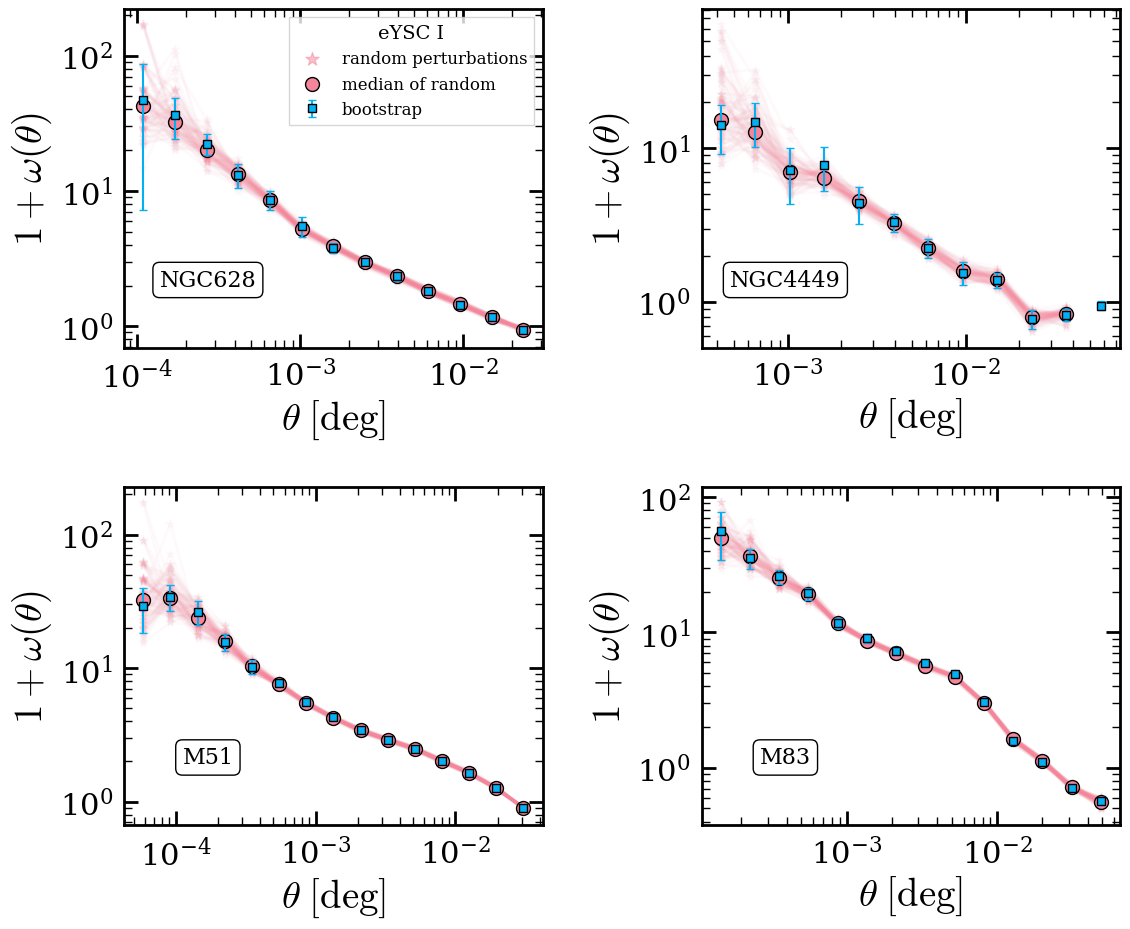}
    \caption{The TPCFs of eYSC Is across the sample with errors calculated via bootstrapping (blue squares) plotted alongside the results of 100 random perturbations to the eYSC I positions (opaque pink stars and lines) and the median of the random perturbation sample (pink circles).}
    \label{fig:pertTest}
\end{figure*}

If random variations in the spatial distribution produce TPCFs aligned with our results with errors calculated via bootstrapping, it follows that bootstrapping provides an adequate estimate of error, in turn validating our fits to the TPCFs. Figure \ref{fig:pertTest} shows the results of this test for the eYSC Is across the sample. It is visually apparent that the points with errors calculated via bootstrapping (i.e., those used for fitting) are in agreement with the range of values calculated via the random perturbation test.

%% For this sample we use BibTeX plus aasjournalv7.bst to generate the

\bibliography{refs}{}
\bibliographystyle{aasjournalv7}



\end{document}

%% file: Table1.tex
\begin{deluxetable*}{cccccccccccc}[bth]
           \tablecaption{An overview of the sample used in this work. From left to right: the name of the galaxy; its morphology as listed in NED; its morphological T-type as listed in Hyperleda; the distance to the galaxy in Mpc, adopted from (in order) \cite{leroy_21}, \cite{legus_main}, \cite{m51dist}, \cite{dellabruna2022}; inclinations of the galaxy in degrees, adopted from (in order) \cite{lang20, hunter05, colombo14, comte81}; standard isophotal radii adopted from \cite{deV91}; galaxy stellar mass adopted from \cite{leroy_21} for NGC 628 and M83, and \cite{legus_main} for NGC 4449 and M51; galaxy total HI mass adopted from the same works as total stellar mass; UV-SFRs adopted from the same works as total stellar mass; the number of eYSCs I, eYSCs II, and YSCs used in the analysis following the sample cuts described in Section \ref{sec:cuts}.
           \label{tab:galaxies}}
\tablecolumns{12}
        \setlength{\extrarowheight}{4pt}
        \tablewidth{\linewidth}
        \tabletypesize{\small}
        \tablehead{
        \colhead{Galaxy} &
        \colhead{Morphology} &  
        \colhead{T-Type} &
        \colhead{$D$} &
        \colhead{$i$} &
        \colhead{$R_{25}$} &
        \colhead{$\log(M_{*}$)} &
        \colhead{$\log(M_{\mathrm{HI}})$} &
        \colhead{SFR} & 
        \colhead{$N_{\mathrm{eYSC\;I}}$} & 
        \colhead{$N_{\mathrm{eYSC\;II}}$} & 
        \colhead{$N_{\mathrm{YSC}}$}
        }
\startdata
    \hline
    & & & [Mpc]& [deg] & [kpc]& [$M_{\odot}$]& [$M_{\odot}$]& $[M_{\odot}\;\mathrm{yr^{-1}}]$ & & &\\
    \hline
NGC628 & SAc & 5.2 & 9.84 & 8.9 & 5.23 & 10.3 & 9.7 & 1.74 & 628 & 487 & 2764 \\
\hline
NGC4449 & IBm & 9.8 & 4.0 & 68.0 & 3.1 & 9.0 & 9.32 & 0.94 & 167 & 182 & 430 \\
\hline
M51 & SAbc & 4.0 & 7.5 & 22.0 & 5.6 & 10.4 & 9.36 & 6.88 & 1554 & 737 & 1890 \\
\hline
M83 & SAB(s)c & 5.0 & 4.7 & 24.0 & 9.18 & 10.5 & 9.98 & 4.17 & 958 & 400 & 2507 \\
\hline
\enddata
\end{deluxetable*}

%% file: Table2.tex
\begin{deluxetable*}{ccccccc}[bth]
    \tablecaption{
        \textnormal{Summaries of the best fit models for each cluster population for each galaxy in our sample. From left to right: the name of the galaxy, the eight cluster populations analysis was performed on, the model which was deemed to best fit that populations TPCF with `PL', `PW', and `T' corresponding to the single, piecewise, and trucnated power law models, respectively, the inner power law slope $\alpha_1$, outer power law slope $\alpha_2$ when applicable, the break scale $\beta$ in units of pc when applicable, and the exponential truncation radius $\theta_c$ in units of pc, when applicable.}
        \label{tab:models}
    }
    \tablecolumns{4}
    \setlength{\extrarowheight}{4pt}
    \tablewidth{\linewidth}
    \tabletypesize{\small}
    \tablehead{
    \colhead{Galaxy} &
    \colhead{Cluster Class} &  
    \colhead{Best-Fit Model} &
    \colhead{$\alpha_1$} & 
    \colhead{$\alpha_2$} &  
    \colhead{$\beta$ [pc]} & 
    \colhead{$\theta_{\rm c}$ [pc]}
    }
\startdata
\hline
NGC628 & \makecell{eYSC I \\ eYSC II \\ eYSC I + II \\ oYSC $\leq$ 10 Myr \\ YSC $>$ 10 yr \\ (0, 10] Myr \\ (10, 100] Myr\\ (100, 300] Myr} & \makecell{PW \\ PW \\ PW \\ PW \\ T \\ PW \\ T \\ T} & \makecell{-1.02 \\ -0.946 \\ -0.921 \\ -0.717 \\ -0.142 \\ -0.761 \\ -0.18 \\ -0.205} & \makecell{-0.519 \\ -0.532 \\ -0.564 \\ -0.389 \\ - \\ -0.38 \\ - \\ -} & \makecell{227.0 \\ 263.0 \\ 281.0 \\ 114.0 \\ - \\ 218.0 \\ - \\ -} & \makecell{- \\ - \\ - \\ - \\ 6830.0 \\ - \\ 9160.0 \\ 6400.0}
\\
\hline
NGC4449 & \makecell{eYSC I \\ eYSC II \\ eYSC I + II \\ oYSC $\leq$ 10 Myr \\ YSC $>$ 10 yr \\ (0, 10] Myr \\ (10, 100] Myr\\ (100, 300] Myr} & \makecell{PL \\ PW \\ PW \\ T \\ PW \\ PW \\ PW \\ PL} & \makecell{-0.669 \\ -0.961 \\ -0.811 \\ -0.544 \\ -0.276 \\ -0.728 \\ -0.401 \\ -0.47} & \makecell{- \\ -0.288 \\ -0.237 \\ - \\ -0.985 \\ -1.01 \\ -0.983 \\ -} & \makecell{- \\ 2050.0 \\ 1910.0 \\ - \\ 424.0 \\ 651.0 \\ 352.0 \\ -} & \makecell{- \\ - \\ - \\ 7190.0 \\ - \\ - \\ - \\ -}
\\
\hline
M51 & \makecell{eYSC I \\ eYSC II \\ eYSC I + II \\ oYSC $\leq$ 10 Myr \\ YSC $>$ 10 yr \\ (0, 10] Myr \\ (10, 100] Myr\\ (100, 300] Myr} & \makecell{PW \\ PW \\ PW \\ PW \\ T \\ PW \\ T \\ T} & \makecell{-0.794 \\ -0.529 \\ -0.694 \\ -0.53 \\ -0.191 \\ -0.49 \\ -0.315 \\ -0.14} & \makecell{-0.505 \\ -0.952 \\ -0.515 \\ -0.807 \\ - \\ -0.68 \\ - \\ -} & \makecell{73.9 \\ 1930.0 \\ 102.0 \\ 927.0 \\ - \\ 1330.0 \\ - \\ -} & \makecell{- \\ - \\ - \\ - \\ 5500.0 \\ - \\ 8590.0 \\ 4770.0}
\\
\hline
M83 & \makecell{eYSC I \\ eYSC II \\ eYSC I + II \\ oYSC $\leq$ 10 Myr \\ YSC $>$ 10 yr \\ (0, 10] Myr \\ (10, 100] Myr\\ (100, 300] Myr} & \makecell{PW \\ PL \\ PW \\ T \\ PW \\ PW \\ PW \\ PW} & \makecell{-0.617 \\ -0.776 \\ -0.671 \\ -0.645 \\ -0.315 \\ -0.546 \\ -0.376 \\ -0.248} & \makecell{-1.01 \\ - \\ -0.932 \\ - \\ -0.904 \\ -0.959 \\ -0.827 \\ -0.98} & \makecell{433.0 \\ - \\ 458.0 \\ - \\ 885.0 \\ 567.0 \\ 727.0 \\ 1380.0} & \makecell{- \\ - \\ - \\ 8290.0 \\ - \\ - \\ - \\ -}
\\
\hline
\enddata
\end{deluxetable*}

%% file: Table3.tex
\begin{deluxetable*}{cccc}[bth]
    \tablecaption{
        \textnormal{Summary of the physical quantities derived from the TPCFs. From left to right: galaxy name, fractal index $D_2$, and the scale at which their is a slope transition in the TPCF $\lcorr$. Error bars are the 16th and 84th percentile errors derived from the posterior distributions of the fits.}
        \label{tab:physicalquants}
    }
    \tablecolumns{4}
    \setlength{\extrarowheight}{4pt}
    \tablewidth{\linewidth}
    \tabletypesize{\small}
    \tablehead{
    \colhead{Galaxy} &
    \colhead{Cluster Class} & 
    \colhead{$D_2$} & 
    \colhead{$l_{\rm corr}$ [pc]}
    }
\startdata
\hline
NGC628 & \makecell{eYSC I \\ eYSC II \\ eYSC I + II \\ oYSC $\leq$ 10 Myr \\ YSC $>$ 10 yr \\ (0, 10] Myr \\ (10, 100] Myr\\ (100, 300] Myr} & \makecell{$0.981^{+0.155}_{-0.176}$ \\ $1.05^{+0.18}_{-0.285}$ \\ $1.08^{+0.0922}_{-0.102}$ \\ $1.28^{+0.347}_{-0.279}$ \\ $1.86^{+0.0208}_{-0.0205}$ \\ $1.24^{+0.312}_{-0.0583}$ \\ $1.82^{+0.0271}_{-0.0272}$ \\ $1.8^{+0.0916}_{-0.0689}$} & \makecell{$227.0^{+115.0}_{-76.5}$ \\ $263.0^{+599.0}_{-133.0}$ \\ $281.0^{+140.0}_{-114.0}$ \\ $114.0^{+3570.0}_{-44.1}$ \\ $< 6830.0^{+669.0}_{-566.0}$ \\ $218.0^{+79.0}_{-196.0}$ \\ $< 9160.0^{+1950.0}_{-1360.0}$ \\ $< 6400.0^{+3830.0}_{-2460.0}$}
\\
\hline
NGC4449 & \makecell{eYSC I \\ eYSC II \\ eYSC I + II \\ oYSC $\leq$ 10 Myr \\ YSC $>$ 10 yr \\ (0, 10] Myr \\ (10, 100] Myr\\ (100, 300] Myr} & \makecell{$1.33^{+0.0417}_{-0.0417}$ \\ $1.04^{+0.0754}_{-0.0629}$ \\ $1.19^{+0.0525}_{-0.0579}$ \\ $1.46^{+0.0931}_{-0.0749}$ \\ $1.72^{+0.122}_{-0.114}$ \\ $1.27^{+0.147}_{-0.0631}$ \\ $1.6^{+0.18}_{-0.157}$ \\ $1.53^{+0.126}_{-0.126}$} & \makecell{$> 2880.0^{+1340.0}_{-1370.0}$ \\ $2050.0^{+526.0}_{-1880.0}$ \\ $1910.0^{+465.0}_{-1870.0}$ \\ $< 7190.0^{+6820.0}_{-3110.0}$ \\ $424.0^{+120.0}_{-87.9}$ \\ $651.0^{+359.0}_{-462.0}$ \\ $352.0^{+149.0}_{-108.0}$ \\ $> 2930.0^{+1340.0}_{-1400.0}$}
\\
\hline
M51 & \makecell{eYSC I \\ eYSC II \\ eYSC I + II \\ oYSC $\leq$ 10 Myr \\ YSC $>$ 10 yr \\ (0, 10] Myr \\ (10, 100] Myr\\ (100, 300] Myr} & \makecell{$1.21^{+0.151}_{-0.139}$ \\ $1.47^{+0.0306}_{-0.0285}$ \\ $1.31^{+0.188}_{-0.0773}$ \\ $1.47^{+0.0397}_{-0.0373}$ \\ $1.81^{+0.0242}_{-0.0242}$ \\ $1.51^{+0.0197}_{-0.168}$ \\ $1.68^{+0.0335}_{-0.0332}$ \\ $1.86^{+0.0597}_{-0.0612}$} & \makecell{$73.9^{+35.1}_{-31.3}$ \\ $1930.0^{+1310.0}_{-1860.0}$ \\ $102.0^{+1730.0}_{-37.4}$ \\ $927.0^{+359.0}_{-254.0}$ \\ $< 5500.0^{+531.0}_{-446.0}$ \\ $1330.0^{+599.0}_{-1310.0}$ \\ $< 8590.0^{+2160.0}_{-1450.0}$ \\ $< 4770.0^{+1220.0}_{-789.0}$}
\\
\hline
M83 & \makecell{eYSC I \\ eYSC II \\ eYSC I + II \\ oYSC $\leq$ 10 Myr \\ YSC $>$ 10 yr \\ (0, 10] Myr \\ (10, 100] Myr\\ (100, 300] Myr} & \makecell{$1.38^{+0.0339}_{-0.0318}$ \\ $1.22^{+0.0201}_{-0.0202}$ \\ $1.33^{+0.0264}_{-0.025}$ \\ $1.36^{+0.0194}_{-0.0196}$ \\ $1.68^{+0.0204}_{-0.0226}$ \\ $1.45^{+0.0157}_{-0.0203}$ \\ $1.62^{+0.0319}_{-0.0315}$ \\ $1.75^{+0.0513}_{-0.0523}$} & \makecell{$433.0^{+53.2}_{-46.8}$ \\ $> 2880.0^{+1370.0}_{-1360.0}$ \\ $458.0^{+93.4}_{-66.2}$ \\ $< 8290.0^{+1720.0}_{-1210.0}$ \\ $885.0^{+62.9}_{-61.8}$ \\ $567.0^{+61.6}_{-44.2}$ \\ $727.0^{+121.0}_{-143.0}$ \\ $1380.0^{+197.0}_{-198.0}$}
\\
\hline
\enddata
\end{deluxetable*}